\documentclass[12 pt]{article}
\usepackage{makecell}
\usepackage{subcaption}
\usepackage{cite}
\usepackage{relsize}
  \usepackage{graphicx}
\usepackage{tikz}
\usepackage{amsmath}
\usepackage{bm}

\newtheorem{theorem}{Theorem}[section]

\def\bi{\begin{itemize}}
\def\ei{\end{itemize}}
\def\bn{\begin{enumerate}}
\def\en{\end{enumerate}}
\def\bq{\begin{eqnarray}}
\def\eq{\end{eqnarray}}
\def\bqn{\begin{eqnarray}}
\def\eqn{\end{eqnarray}}
\def\bqta{\begin{eqtarraya}}
\def\eqta{\end{eqtarraya}}
\def\bqtb{\begin{eqtarrayb}}
\def\eqtb{\end{eqtarrayb}}
\def\bqtc{\begin{eqtarrayc}}
\def\eqtc{\end{eqtarrayc}}
\def\be{\begin{equation}}
\def\ee{\end{equation}}

\def\bea{\begin{eqnarray}}
\def\eea{\end{eqnarray}}
\def\beann{\begin{eqnarray*}}
	\def\eeann{\end{eqnarray*}}

\def\bmat{\left[ \begin{array}}
	\def\emat{\end{array} \right]}

\def\bsmat{\left[ \begin{smallmatrix}}
	\def\esmat{\end{smallmatrix} \right]}

\def\red{\textcolor{black}}
\def\cyan{\textcolor{black}}

\begin{document}

\title{LIDAR-Assisted Exact Output Regulation for Load Mitigation in Wind Turbines }

\author{Amin~Mahdizadeh,        Robert~Schmid
        and~Denny~Oetomo% <-this % stops a space
\thanks{A. Mahdizadeh and  R. Schmid are with the Department
of Electrical and Electronic Engineering, University of Melbourne, 3010 Australia.
D. Oetomo  is with the Department
of Mechanical Engineering, University of Melbourne, 3010 Australia.
  email: amahdizadeh@student.unimelb.edu.au, rschmid@unimelb.edu.au, doetomo@unimelb.edu.au}
  % <-this % stops a space
}

\maketitle
\begin{abstract}

Optimising wind turbine performance  involves maximising  energy  harvesting   while seeking to  minimise load fatigues  on the
tower structure,  blades and rotor.  The problem is  inherently  difficult   due to the slow response of wind turbines compared to  wind variation frequencies.  To improve turbine control performance,  wind preview measurement  technologies such as Light Detection And Ranging (LIDAR)  have  been a point of interest for researchers in recent years. \cyan{However,} the effective augmentation of  wind preview  information into control methodologies has to  date  proven  to be a challenging  problem.  In this paper,  we  explore the application of  a classical  control  methodology known  as Exact Output Regulation (EOR) for  improving the  control performance of a LIDAR-enhanced  wind turbine.  The EOR controller is designed to achieve the  rejection of  known  input disturbances, while also ensuring the system output tracks  a desired  reference signal.   The controller is comprised of a state feedback  controller together with a feedforward  gain.  The LIDAR  wind  preview information is used to  obtain  a low-order  exosystem for modeling  wind dynamics.   This  wind exosystem is  used to obtain  the feedforward gain matrix that  enables the EOR controller  to  effectively reject the input  disturbance and achieve the desired reference tracking.
Extensive simulations of the EOR controller  with  a broad range of wind speeds  in both partial load and full load operating  regions  are performed on the full nonlinear aero-elastic model of the National Renewable Energy Laboratory (NREL) 5 MW reference wind turbine.
For performance comparisons, we also implement a  Baseline  torque controller  and a commonly used feedforward control method known as Disturbance Accommodation Control (DAC).    The results show that, in comparison with a Baseline and DAC controller, the EOR controller can provide substantially  improved reduction of  fatigue  loads and smoother power output,  without compromising  energy production levels.

\end{abstract}

%\begin{IEEEkeywords}
%Fatigue Load Mitigation in Wind Turbines, Exact output regulation (EOR), feedforward control, Light detection and ranging (LIDAR).
%\end{IEEEkeywords}

% \fi
%\IEEEpeerreviewmaketitle

\section{Introduction}

Reliable power production from wind is a difficult problem,   due to the intermittent nature of the wind. It  has been the subject of research interest  from the early days of electrical wind turbines.  The transformation of the  free kinetic energy of the wind into mechanical, and subsequently, electrical energy comes with the cost of the structure and materials of the turbine as well their maintenance.
  Two simultaneous approaches  taken  to  reduce  the  levelised cost of wind energy (LCOE)  \cite{chen2017wind}  are  maximizing the energy harvesting efficiency  and reducing the cost of maintenance by reducing  turbine fatigue  loads.   From the control engineering point of view, performance improvements  may be achieved through a combination of additional 
  measurements and superior  control methodologies.

Mechanical loads on the wind turbine structure induced by sudden variations of the wind  can be mitigated by the control of the blade pitch angle  and  generator torque  control \cite{muljadi2001pitch}. Energy harvesting efficiency can be improved by more precisely steering the wind turbine states on their optimal trajectories.
The use of feedback control is the conventional approach for  stabilizing and regulating dynamical systems, however,    feedback control for turbines
may not yield  satisfactory system behaviour,  as it assumes the turbine only reacts to the variations of the wind which may have already influenced the system states and driven them away from their desired values. To address this shortcoming, Light Detection And Ranging (LIDAR) has been proposed as a new technology to provide estimates  of upcoming wind speeds prior to the wind interacting with the turbine blades \cite{harris2006lidar}.   Recent LIDAR cost  reductions  have  opened a new research area on the  use of feedforward control  for large scale wind turbines,  using nacelle-based  LIDAR systems to obtain real-time wind speed and direction information up to several hundred meters ahead of the wind turbine \cite{wang2013comparison,scholbrock2016lidar}.  \red{Recently  \cite{goit2019lidar} collected extensive  field data which showed that profiling \cyan{LIDARs} can measure wind speeds and turbulence intensities with acceptable accuracy,  in \cyan{comparison} with  tower-mounted cup and sonic anemometers. }

An early work involving LIDAR for feedforward control of the wind turbines \cite{schlipf2008prospects} showed that an augmented LIDAR feedforward control may improve  turbine energy harvesting and load reduction. Further investigations  using non-causal series expansion \red{model-inverse} method appeared in \cite{dunne2011adding},  where  it was shown that  a lower damage equivalent load (DEL) on tower root fore-aft oscillations could be obtained relative to a Baseline controller, with no loss in the produced power. In \cite{laks2010blade}, a preview-based feedforward method assuming highly idealized wind measurements  concluded that wind evolutions in more realistic conditions can eliminate advantages gained by using preview-based feedforward techniques.  The LIDAR-assisted design in 
 \cite{dunne2010combining} used three different model inversion methods  to augment  the feedback loop: the nonminimum-phase zeros ignore (NPZ-Ignore), the zero-phase-error tracking controller (ZPETC) and the zero-magnitude-error tracking controller (ZMETC); some improvement in some of the loads were obtained.
%More realistic wind measurements were considered in \cite{schlipf2010look}  where  a robust feedforward controller using a LIDAR simulator was presented.
Two  early field testing surveys of LIDAR-based feedforward control  using model inversion methods  were carried out in \cite{scholbrock2013field,schlipf2014field}.  The results showed evidence of tower load reduction by $10 \%$ due to the utilization of LIDAR, confirming the previous results on simulations. In \cite{haizmann2015optimization} load reductions were improved by using Continuous-Wave LIDAR. Although model inversion methods are feasible in the presence of look-ahead LIDAR information, they require the use of approximated models of the plant inverse to avoid the effects of non-minimum phase zero inversion.

LIDAR-assisted control has also been tested for improving energy harvesting at below rated wind speeds. Results from \cite{schlipf2011prospects}  showed that LIDAR-aided rotor speed and yaw angle control yielded increased energy production. Field tests of the methods proposed in\cite{schlipf2011prospects} were  extended in \cite{schlipf2013direct} with real data collected from LIDAR where $0.3 \%$ improvement in energy gain was achieved, however this  came at the cost of doubling the  loads on the shaft.
In \cite{wang2013comparison} three different methods of wind turbine control are augmented by LIDAR and compared at \cyan{below-rated} wind speeds: the optimally tracking rotor (OTR) control scheme \cite{johnson2006control}, Preview Control \cite{takaba2003tutorial} and Disturbance Accommodation Control (DAC),  sometimes also known as Disturbance Tracking Control (DTC) \cite{stol2003disturbance}. However,  these methods were only able to  increase energy harvesting by very small amounts, and these improvements came at the cost of substantial increases  in some  fatigue loads.

DAC has been one of the most widely used feedforward methods for  wind turbine control during the last decade,  due to its simplicity and  capacity to estimate the effective wind speed on the rotor. It  was first applied to  wind turbines in  \cite{balas1998disturbance}, to counteract the effects of wind disturbances. Later, DAC methods have been used in \cite{wright2004modern} and  \cite{hand2004advanced}  to reduce blade fatigue loads induced by the wind disturbances. Also in \cite{stol2003periodic} it was applied for canceling asymmetric blade mass effects of a two-bladed wind turbine causing periodic loads. A field test on  the  Controls Advanced Research Turbine (CART) \cite{wright2006testing} showed that, compared to a Baseline PI controller, DAC can reduce structural dynamic loads in real case scenarios.

Efforts have  been made to  enhance DAC by  augmenting the controller with  LIDAR information, replacing the estimated wind speed with the LIDAR measured speed. However, to  date  only  modest performance improvements have  been achieved.  \cite{wang2013comparison} showed that the LIDAR augmented DAC (known as DAC+LIDAR) achieved less than $0.5 \%$ improvement in power production,  and this improvement  came at  the cost of a $5.7 \%$ increase in rotor fatigue  load.
% Another improved attempt to use DAC with LIDAR wind information was carried out in \cite{wang2014lidar} where the %structural loads were decreased by $2 \%$.

Model Predictive Control (MPC) methods have also been a point of interest since they can  readily accommodate  LIDAR wind information.  \red{The simulation studies \cite{schlipf2013nonlinear,tofighi2015nonlinear} showed that nonlinear MPC can reduce structural loads for turbulent winds and  extreme  loads gusts.} However, a  key limitation of nonlinear  MPC is the substantial computational cost, making it infeasible for real-time applications on conventional industry grade controllers.

In this paper, we present a novel method for turbine control that can make effective  use of  LIDAR information, without requiring the excessive  computational costs of MPC. Our  approach will employ  the  Exact Output Regulation (EOR)   control methodology,  and  our results will  show that it can deliver substantial  reductions  in  fatigue loads,  without compromising energy  harvesting.  Additionally, its rapid computation  time makes it  feasible for real-time implementation \cite{mahdizadeh2018comparison}.

The EOR control methodology has played a central  role in modern control  systems design for several  decades \cite{trentelman2002control}-\cite{saberi2012control}.   The output regulation problem considers a linear time invariant (LTI) plant that is assumed to be subject to known input time-varying disturbances, and whose output is desired to track a known time-varying reference signal. The reference signals and external disturbances are modelled as the  outputs of a  linear  exosystem.  Solution of the problem  requires the design  a combined  state feedback and feedforward  controller that will internally stabilise the plant  while rejecting the disturbances and ensuring the output converges asymptotically to the desired reference signal.    The required feedback  and feedforward gain matrices  are readily computable \cite{saberi2012control}.

For turbine  control,  effective disturbance rejection involves the minimisation of wind  disturbances  on the control  actuation. These are the  rotor  torque and also  the blade  pitch angle.  Effective reference tracking involves operation of the turbine rotor and  blades so  as to  generate the  optimal energy from  the  wind.  The problem is difficult as  the  wind frequency variation  (turbulence) is  much faster than the turbine response.
The availability of LIDAR wind preview information enables the  wind signal  to be  modeled as  a low-order  linear  dynamical  system.  This linear  system may then be  incorporated into the  EOR  control methodology as an exosystem whose outputs provide the input disturbance and the output reference signal \cite{mahdizadeh2017enhanced,mahdizadeh2017fatigue}.
 When  the turbine is  operating below its rated power, the reference  signal is the value of the rotor speed  that achieves the maximum power generation. When the turbine operates at its rated power, the reference is the value of the blade pitch angle that maintains the  turbine  at its  rated power.

A  simulation  environment known as Turbine Output Regulation (TOR) has been developed by the authors to apply EOR to the  control of  a  5 $MW$ reference Horizontal Axis Wind Turbine (HAWT) \cite{jonkman2009definition}. To  simulate the turbine response,  the  Fatigue, Aerodynamics, Structures,
and Turbulence (FAST)  code   using a high-order aero-elastic nonlinear  turbine model developed  by  National Renewable Energy Laboratory (NREL) \cite{jonkman2005fast} will be  used. The NREL TurbSim package \cite{jonkman2009TurbSim}  is used to simulate realistic wind fields,  and damage equivalent loads (DELs) are computed by  the Rain-Flow-Counting-Algorithm open source MATLAB$^{\textregistered}$ code \cite{RFCToolbox}.   For performance comparison purposes, TOR obtains a feedforward controller using  the DAC method,  and also  a Baseline torque controller. Energy  harvesting  and  DELs are computed within the  TOR simulation  environment to  allow performance comparisons of these three controllers, for a broad range of wind speeds and  intensities.

Our simulations studies on the 5 $MW$ HAWT model show that, in comparison with DAC and the Baseline  method,  the EOR controller can substantially reduce the fatigue loads on the tower,
blades and low speed shaft.  Additionally, EOR is able to reduce the  standard deviation of rotor speed and output power,  without any  loss in  energy harvesting.  The authors believe that  EOR is able to obtain  these  improvements  through its modeling of the wind  dynamics.  Where DAC  treats the wind signal as a constant disturbance and does not   consider any output tracking objective, EOR is able to accommodate derivatives  of the wind signal, leading to improved tracking performance and  disturbance rejection.

The remainder of this paper is organized as follows. Section \ref{sec:Wmod} describes  the turbine modelling  in both rated  and below rated operating  regions.
Section  \ref{sec:WCon}  introduces the control performance objectives and  methods to  be considered in our simulation  studies.  We provide a summary  of the EOR control methodology for a general  linear time  invariant plant in  state space form.  The Baseline feedback controller and DAC control method   are also  described.  In Section \ref{EOR_WT},  we consider how to  use raw LIDAR wind measurement data to  develop a suitable  low-order linear   wind  model  that will be  used as the   exosystem   for the development of an EOR controller for the turbine.
Section \ref{sec:SimEnv} describes the turbine simulation environment   used  for our  performance  comparisons,  and our simulation results will be presented  and   discussed  in Section \ref{sec:SimComp}.  Finally, in section \ref{sec:Conc} our  conclusions and future work will  be presented.

%%%%%%%%%%%%%%%%%%%%%%%%%%%%%%%%%%%%%%%%%%%%%%%%%%%%%%%%%%%%%%%%%%%%%%%%%%%%%%%%%%%%%%%%%%%%%%%%%%%%%%%%%%%%%%%%%%%%%%%%%
\section{Wind Turbine Modeling }
\label{sec:Wmod}

The wind turbine model used in this work is the NREL 5-MW reference Horizontal Axis Wind Turbine (HAWT)  \cite{jonkman2009definition}.  Detailed specifications of the  turbine are  presented in  Table \ref{table:5MW}.
%%
%     Kd = 867e6;
%Cd = 6.2e6;
%
%J_rot = 11776047;
%J_gen = 534*N^2;

\begin{table}
	\caption{The NREL 5-MW Wind Turbine Specifications}
	\label{table:5MW}
	\centering	
	\begin{tabular}{p{4cm} c r}
	      \red{Quantity}   & \red{Symbol}  & \red{Value} \\
		\hline \hline
		Rated power & $ \red{P_{rated}}  $ & 5 $MW$ \\
		Rated rotor speed & $ \red{\Omega_{rated}}$  & 12.1 $rpm$ \\
		Rated wind speed & $\red{V_{rated}} $   & 11.4 $m/s$ \\
		\red{Rated generator  torque }& \red{$M_{rated}$}   & \red{ 43.1   $ kNm $}  \\
		Cut-in wind speed & $V_{in}$ & 3 $m/s$ \\
		Cut-out wind speed & $V_{out}$ & 25 $m/s$ \\ % \hline
		Rotor radius & $R$ & 63 $m$ \\
		Hub height & $h_H$ & 90 $m$ \\
		Rotor moment of inertia & $J_{r}$ &   11.8 $ kt/{m^2}$ \\
		Generator moment of inertia & $J_{g}$ & 534  $kg/{m^2}$ \\
		Drive-train Stiffness & $K_d$ & $867 $ $MNm/rad$ \\
		Drive-train Damping & $C_d$ & $6.2 \times 10^6 $  \\
		
		%Total moment of inertia about low speed shaft & $J$ & $11.77 \times 10^6$ $kg/{m^2}$  \\
		Gearbox ratio & $i$ & 1/97 \\
		Tower equivalent modal mass & $m_{Te}$ & $4.36 \times 10^3 kg$ \\
		Tower structural damping & $c_{Te}$ & 17782  \\
		Bending stiffness & $k_{Te}$ & $1.81 MN/m$  \\
		Static tower-top displacement in absence of thrust forces & $x_{T0}$ & -0.0140 $m$ \\
		%		Natural frequency of first tower fore-aft bending & $f_0$ & 0.32 $Hz$ \\
		%		Structural damping ratio & $d_s$ & 0.01 \\
		Blade pitch actuator undamped natural frequency & $\omega$ & $2\pi$ $rad/s$  \\
		Blade pitch actuator  damping factor & $\zeta$  & 0.70 \\  % \hline
		Optimal tip speed ratio &\red{ $\lambda^*   $} & 7.55 \\
		Maximum power coefficient  &\red{ $C_{p}^*$ }  & 0.482 \\
		%		Maximum generator torque rate & $\dot{M}_{g,\rm max}$ & 15000 $N.m/s$ \\
		\hline
	\end{tabular}
\end{table}

\subsection{Operating  Regions}

\begin{figure}[t]
\centering
	\includegraphics[width=9cm]{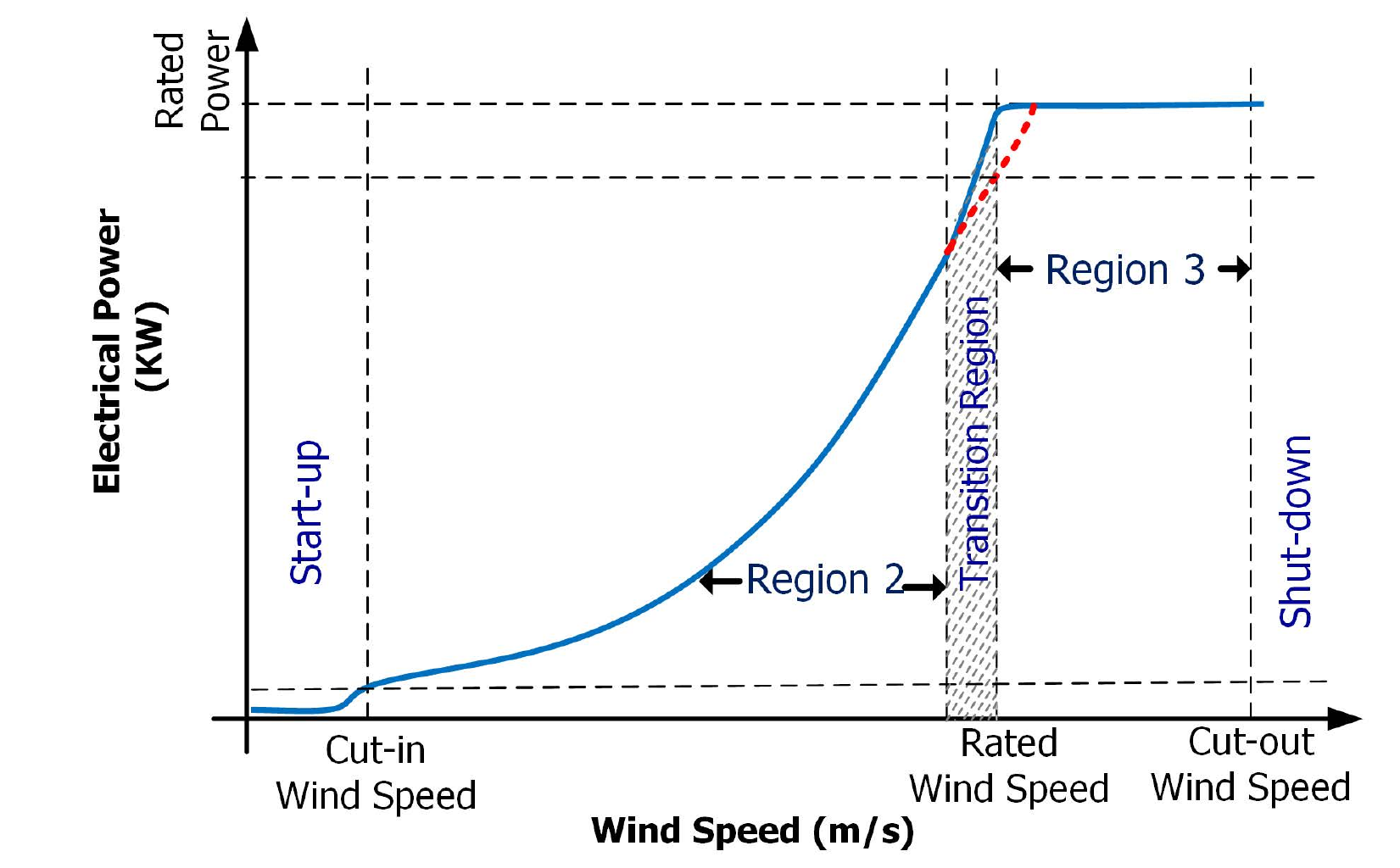}
	\caption{Wind Turbine Operating Regions \cite{tofighi2015nonlinear} }
	\label{fig:regions}
\end{figure}

Turbine  operation may be divided into four distinct regions,   determined by the  mean wind speed,  as  depicted in Figure \ref{fig:regions}.   The  \emph{start-up} region  applies   for  very low wind speeds,  where the  kinetic energy of wind is  insufficient  for turbine operation.  At the  \emph{cut-in} wind speed (about  3 $ m/s$), the turbine begins to operate  in \emph{Region 2}.  The higher the wind speed, the   greater the energy that can be harvested by the blades.  The \emph{rated power}  of the turbine is determined by  factors such as the mechanical load capacity of the components as well as the limits on the electrical power and currents deliverable by the generator. The wind speed at which the turbine is  able to  generate  its  rated power  is called the \emph{rated wind speed},  (about  11.4 $ m/s$), and wind speeds above this are
 referred to as \emph{Region 3}.  The \emph{shut-down}  region applies when the wind speed exceeds a safe limit known as the  \emph{cut-out} wind speed (about 25 $m/s$).
 %Power generation is shut down and the turbine is placed in a safe position.

\subsection{Optimal Power Generation} \label{SecOPG}

The {\it instantaneous power}  carried by the  air moving  through the  vertical plane of the rotor blades is given by
\begin{equation}
\label{eq:Pw}
P_{wind} = \frac{1}{2} \rho A_R v_{\texttt x}^3,
\end{equation}
where     $A_R$ is the swept  area  of the blades with  radius $R$,  $\rho$ is the air density and   $v_{\texttt x}$ is the magnitude of the  component of the wind velocity vector that is perpendicular to the rotor plane. We refer to  this perpendicular component as the  {\it   longitudinal wind speed}. \red{The \textit{mechanical power} extracted from the wind  is  $P_{mech} = \Omega_r M_r$,  where $\Omega_r$ 
and   $M_r$  are the rotor speed and rotor torque, respectively.}

The {\em blade tip speed ratio} (TSR) is the ratio of  the linear speed of the blade tip to the longitudinal wind speed  $v_{\texttt x}$ that can be formulated as:
\begin{equation}
\label{eq:tsr}
\lambda(\Omega_r , v_{\texttt x}) := \frac{\Omega_r R}{v_{\texttt x}}.
\end{equation}

  This efficiency of the conversion of the wind energy  to rotational-mechanical energy by the blades is the    \emph{power coefficient}  of  the turbine and is defined as
% to be the ratio of the extractable power  to the available wind energy:
\begin{equation}
\label{eq:C_p}
C_p(\lambda , \theta) := \frac{P_{mech}}{P_{wind}},
\end{equation}
\red{where $\theta$ is the  blade  pitch angle}. 
 According to  Betz' law \cite{van2007lanchester}, the power coefficient is  upper bounded 59.3 \%.  However, most wind turbines  in real operating conditions fall short of this limit.
The power coefficient  is  provided as a 2-D lookup table by the  turbine manufacturer.

The {\it aerodynamic torque of the rotor}  is modeled by the equation
\begin{equation} % M_a
\label{eq:M_a}
M_a(\Omega_r, v_{\texttt x},\theta) := \frac{1}{2}\rho \pi R^3\frac{C_p(\lambda, \theta)}{\lambda}v_{\texttt x}^2,
\end{equation}

In Region 2,   the main objective of the turbine controller  is to convert as much of the available  wind energy into mechanical energy as  possible.   To achieve this, the  blade pitch angle is kept at $\theta = 0$ and  a rotor torque controller is  used to maintain the TSR  at the optimal  value $\lambda^{*}$ that maximizes $C_p$; thus
%  According to Table \ref{table:5MW},  for the 5 MW HAWT  turbine,  $\lambda^{*}=7.55$,  and we  have
\begin{equation}
\label{cpmax}
C_p^*   = C_p(\lambda^{*} , 0).
\end{equation}

Using \eqref{eq:tsr},  \eqref{cpmax}   and $\lambda^{*}$, we obtain $\Omega_r^*$,
 the  rotor speed  that  yields  optimal energy harvesting. 

In Region  3,  the convertible power $P_{mech}$ is  larger than the wind turbine's rated power.  Hence the maximum power extraction objective no longer applies,  and the control objective becomes that of maintaining the turbine power output  constant  at the  rated  power level.  This  is achieved by   keeping the generator  torque at the  rated value,  and the  blade pitch angle $\theta^*$  is chosen  to  ensure the
  aerodynamical torque of the rotor is at its rated value.
 % Here  $M_{rated} = P_{rated} / \Omega_{rated}$ is the rated aerodynamical torque of the rotor.

% In another word, the main control objective in this region is to keep the wind turbine states such as rotor speed in the nominal value meanwhile reducing the mechanical fatigue and extreme loads induced by the variations of turbulent wind.

\subsection{Nonlinear Wind Turbine Model}
Wind turbines are nonlinear systems consisting of several dynamic components coupled together.
In our simulations, we use  the open source  NREL FAST  7 code \cite{jonkman2005fast} to provide a high-fidelity turbine response simulation. However, a lower (reduced) order model is required for the model-based controller design.
The  fifth-order nonlinear model from \cite{mirzaei2012wind} describes the turbine drive-train as a two-mass system,  and
the blade pitch actuation system is modeled by   a second order linear system with   damping parameter $\zeta$ and natural frequency $\omega$ according to Table \ref{table:5MW}. This model is described as:

%\begin{equation} % 
%\label{eq:torsion}
%\dot{\phi} = \Omega_r - \Omega_g
%\end{equation}
%\begin{equation} % shaftspeed
%\label{eq:shaftspeed}
%J_g \dot{\Omega}_g = C_d(\Omega_R - \Omega_g) + K_d \phi - M_g
%\end{equation}
%\begin{equation} % towerdisp
%\label{eq:genspeed}
%J_r \dot{\Omega}_r = M_a(.) - C_d(\Omega_r - \Omega_g) + K_d \phi
%\end{equation}
%\begin{equation} % pitchservo
%\label{eq:pitchservo}
%\ddot{\theta} + 2\zeta\omega\dot{\theta}= \omega^2(\theta_c - \theta).
%\end{equation}

\begin{eqnarray}\label{eq:1}
J_r \dot{\Omega}_r &= & - C_d(\Omega_r - \Omega_g) - K_d \phi + M_a(\Omega_r, v_{\texttt x},\theta),  \label{eq:genspeed} \\
\dot{\phi} &=  &\Omega_r - \Omega_g  \label{eq:torsion},\\
J_g \dot{\Omega}_g &= &C_d(\Omega_r - \Omega_g) + K_d \phi - M_g, \label{eq:shaftspeed} \\
\ddot{\theta} &= & -  2\zeta\omega\dot{\theta}+\omega^2(\theta_c - \theta), \label{eq:pitchservo}
\end{eqnarray}
where $\Omega_r$,  $\phi$,  $\Omega_g$,  and $\theta$ are the  \red{rotor  speed,  drive train torsion, generator speed and blade pitch angle,} respectively. The generator torque $M_g$ and blade pitch command $\theta_c$ are the control inputs. Parameters $J_r$ and $J_g$ are the moments of inertia of the rotor and generator while  $C_d$ and $K_d$ are the damping and stiffness coefficients of the drive train. It should be noted that in this work  $\Omega_g$ has been normalized for the gearbox ratio \red{$i$}  so that it is in the same range as $\Omega_r$.

\subsection{{Linearized Wind Turbine Model}}
\label{ssec:LWTm}

For the EOR and DAC controller design methods used in this study,  a reduced order linearized model will  be required.   We follow the guidelines for   turbine  model linearization  given in  \cite{wright2004modern}  and
 introduce   $x = [\Omega_r \quad  \phi \quad  \Omega_g \quad \theta \quad \dot{\theta}]^T$ as the state variable vector.  Also $u = [\theta_c \quad M_g]^T$ is the control input vector,  and $d=v_{\texttt x}$  is the input disturbance. 	
 		For a given mean wind speed, $v_{\texttt{x},0} $,  an equilibrium point  $(x^*,u^*,d^*)$ may  be found which satisfies
		\begin{equation}
			\label{eq:eql_point}
			\begin{aligned}
			\dot{x} = f(x^*,u^*,d^*) = 0, \quad
			\text{subject to}:
			\Omega_r = \Omega_r^*  ,
			\end{aligned}
		\end{equation}
		where $\dot x = f(\cdot)$ describes the nonlinear dynamics (\ref{eq:genspeed})-(\ref{eq:pitchservo}).
					For  mean  wind speeds in Region  2,  we have
					\be  \Omega_r^* =\frac{\lambda^*}{ R} v_{\texttt{x},0}  \ee 					
		In Region 3,   $\Omega_r^* = \Omega_{rated}$,   as given in Table \ref{table:5MW}. 
		
	\red{Obtaining Jacobi matrices at the equilibrium point, we obtain the  linear  state space model in  homogenised  coordinates
\begin{equation}  %fullmodel
\Sigma:\; \left\{  \begin{array}{ccl}
\dot{\bar{x}}(t) & = &A  \bar x(t) + B \bar{u}(t) + H \bar{d}(t),  \\
\bar{y}(t) & = &C_y \bar{x}(t),             \\
\bar{z}(t) & = & C_z \bar{x}(t),
\end{array}\right. \label{eq:Sigma}
\end{equation}
where $\bar{x} = x - x^*$, $\bar{ u} = u - u^*$ and $\bar{d} = d- d^*$  represent coordinates homogenised to the equilibrium point.  The homogenised measured output is  denoted as  $\bar y$,  and $\bar z$ is  the homogenised controlled output. }	
%
%where the state vectors and matrices are determined by \eqref{eq:eql_point}-\eqref{CzReg2} , according to  applicable  mean wind %speed and region of operation. 	
		 The state matrices are
\be	A =
	\bmat{ccccc}
	\frac{(\gamma - C_d)}{J_{r}} & \frac{-K_d}{J_{r}} & \frac{C_d}{J_{r}} & \frac{\beta}{J_{r}} & 0               \\
	1                            & 0                  & -1                & 0                   & 0               \\
	\frac{C_d}{J{g}}             & \frac{K_d}{J{g}}   & \frac{-C_d}{J{g}} & 0                   & 0               \\
	0                            & 0                  & 0                 & 0                   & 1               \\
	0                            & 0                  & 0                 & -\omega^2           & -2 \zeta \omega
	\emat,  \label{Adef}
\ee
\be
B =
\bmat{cc}
0 & 0  \\
0 & 0  \\
0 &\frac{ -1}{J_g} \\
0 & 0 \\
\omega^2 & 0
\emat, \
H =
\bmat{c}
\frac{\alpha}{ J_r}  \\
0  \\
0	\\
0	\\
0
\emat, \label{BHdef}
\ee
where
\[
\gamma =\left. \frac{\partial M_a}{\partial \Omega_r} \right|_{x^*,u^*,d^*}, \  \alpha = \left.\frac{\partial M_a}{\partial v_{\texttt x}}\right|_{x^*,u^*,d^*}, \   \beta =\left. \frac{\partial M_a}{\partial \theta}\right|_{x^*,u^*,d^*}
\]
We assume that only   $\Omega_r $,  $ \Omega_g $  and $ \theta$  are  measurable. Therefore, the measurement output matrix $C_y$ will be
\be
C_y = \bmat{ccccc}
1 & 0 & 0 & 0 & 0 \\
0 & 0 & 1 & 0 & 0 \\
0 & 0 & 0 & 1 & 0
\emat.   \label{CyReg3}
\ee
In Region  3 the generator torque $M_g$ is kept on its rated value and the blade pitch angle $\theta^*$
  is  	\red{  obtained by solving
  \be
    M_a(\Omega_{rated}, v_{\texttt{x},0}, \theta^*) = M_{rated}  \label{theta*}
  \ee}
Thus the  input matrix $B$ and controlled output matrix $C_z$ simplify to:
\be
B = [0 \quad 0 \quad  0 \quad 0 \quad \omega^2]^T, \quad C_z = [0 \quad 0 \quad  0 \quad 1 \quad 0].   \label{CzReg3}
\ee

  In  Region 2, the blade pitch angle $\theta$  is maintained at zero,  so  (\ref{eq:pitchservo})  is  not used,  and we
  obtain  a  third order model with the  state variables
\[
x = [\Omega_r \quad  \phi \quad  \Omega_g]^T.
\]
 The control input is the generator torque,  $u(t) = M_g  $.   The   linearized system \eqref{eq:Sigma}  has  matrices
\[
A =
\bmat{ccc}
\frac{(\gamma - C_d)}{J_{r}} & \frac{-K_d}{J_{r}} & \frac{C_d}{J_{r}} \\
1                                         & 0                       & -1                     \\
\frac{C_d}{J_{g}}                    & \frac{K_d}{J_{g}}   & \frac{-C_d}{J_{g}} \\
\emat,
B =
\bmat{c}
0  \\
0  \\
\frac{-1}{J_{g}}
\emat,
H =
\bmat{c}
\frac{\alpha}{J_r}  \\
0  \\
0
\emat
\]
The measurement output matrix reduces to
\be
C_y = \bmat{ccc}
1 & 0 & 0 \\
0 & 0 & 1
\emat .   \label{CyReg2}
\ee
The rotor  speed $\Omega_r$  is the controlled output, thus $C_z$ is
\be
C_z = \bmat{ccc}
1 & 0 & 0
\emat .  \label{CzReg2}
\ee

%%%%%%%%%%%%%%%%%%%%%%%%%%%%%%%%%%%%%%%%%%%%%%%%%%%%%%%%%%%%%%%%%%%%%%%%%%%%%%%%%%%%%%%%%%%%%%%%%%%%%%%%%%%%%%%%%%%%%%%%%%%%%%%%%%%%%%%
%%%%%%%%%%%%%%%%%%%%%%%%%%%%%%%%%%%%%%%%%%%%%%%%%%%%%%%%%%%%%%%%%%%%%%%%%%%%%%%%%%%%%%%%%%%%%%%%%%%%%%%%%%%%%%%%%%%%%%%%%%%%%%%%%%%%%%%%%%%%

\section{Wind Turbine Control }
\label{sec:WCon}

Here  we introduce  the control objectives and  methodologies to  be considered in our simulation studies. We  discuss  the  specific   performance objectives of turbine control, and articulate some  measures for   comparing
 the   performance  of \cyan{different} control methodologies. Lastly, we  introduce   the   Exact Output Regulation control   methodology that  has  been widely studied in the control  systems  literature for several decades.   The principal  novelty of our work  lies in  the  application of this classical  control methodology to  a  LIDAR-enhanced  wind turbine.

\subsection{Wind Turbine Control Objectives}

Wind turbine control objectives  may  be divided into two categories:  improving  power production and  reducing  load fatigues.  In Region  2 the  power objective  is to generate the  maximum  power from the available wind,  while in Region 3, the objective is to maintain the  power at the turbine's rated value. The performance metric for these power objectives are the mean and standard deviation of the associated signal.  For example, the mean value for the generated power will be calculated with the following equation: 
\be
P_{mean} = \frac{1}{T} \int_{0}^{T} P \, dt,
\ee
where $T$ is the time duration of measurement and $P$ is the generator output power measured by FAST.
A  smaller standard deviation of the rotor speed $std(\Omega_r)$ indicates improved performance of the controller on retaining the rotor speed at the rated speed in Region 3.  A  smaller value for the standard deviation of the generated power $std(P)$ indicates reduced  frequency fluctuations of the  power supplied to  the power network;   such  fluctuations are known to be a  problematic aspect of the  injection of  intermittent active power from large wind farms  into the  network \cite{ono2012frequency, yamashita2011development}.  In Region 2, a smaller value for  the standard deviation of the tip speed ratio $ std(\lambda) $ indicates the controller is  more successful in maintaining the tip speed ratio at the optimal value for power generation.

Fatigue loads on the structure are caused by the oscillations induced by the wind and actuator variations. The standard metrics for measuring turbine fatigues are the Damage Equivalent Loads (DELs)  which represent the damages caused by the structural loads accumulated during the lifetime of the  turbine.    The three principal loads considered in this paper
%for calculating DELs which
are the torsional  displacement  on the drive train shaft, and the  bending moments  of  the tower root and blade root. These points are under high strain and also compose the most expensive parts of the wind turbine. Therefore, load mitigation on these points are economically very  desirable. Finally, for visualization in the frequency domain\cite{schlipf2013nonlinear}, power spectral density (PSD) of some of the measurements will be shown.  PSD represents the   spectral  content of these signals. Normally,  reduced  high  frequency spectral content is desirable as it implies reduced vibrations in the system  components.

%Moreover, commanded generator command torque rate (CTR ) will be measured to determine which controller causes less actuation on the outputs which represents the control effort being used for achieving other control objectives. Finally, since the effect of using different controllers along with LIDAR is more suitable for visualization in the frequency domain\cite{schlipf2013nonlinear}, power spectral density (PSD) of some of the measurements will be shown. PSD represents how different controllers behave in different frequencies. Normally, higher attenuation in high frequencies are desirable as it represents less vibrations on the actuators.

\subsection{The Exact Output Regulation Control Methodology}\label{secEOR}

EOR is a multi-variable LTI control methodology in which the plant is subject to known time-varying input  disturbances that are to be rejected, and the  plant outputs are required to track a known time-varying reference signal. The aim of EOR is to design a feedback control law which ensures that the plant dynamics are stable, and
the output asymptotically converges to a desired reference signal while rejecting the disturbances.    The  following summary of EOR is  taken  from \cite{schmid2014nonovershooting},  which   was adapted from \cite{saberi2012control}.

\begin{figure}[t]
	\centering
	\includegraphics[trim={0 0 0 0},clip, width=.5\textwidth]{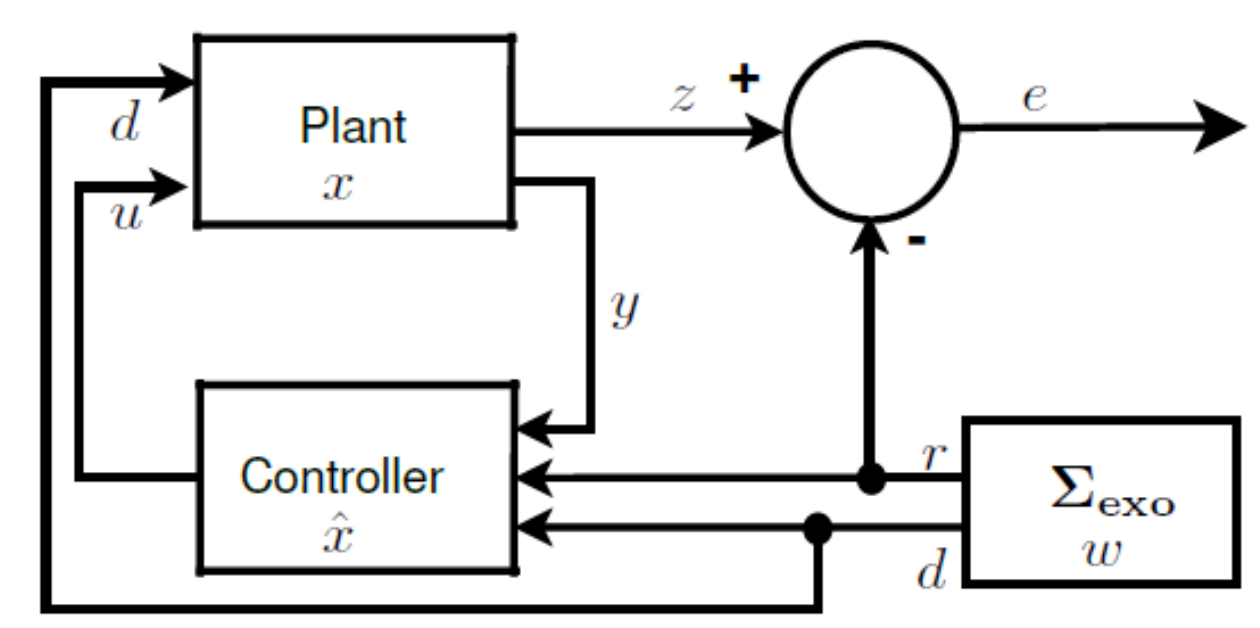}
	\caption{Block diagram of EOR control methodology}
		\label{fig:EOR}
\end{figure}

The EOR control methodology considers a linear time-invariant  multivariable system  shown in  the block  diagram of Figure \ref{fig:EOR}.  The plant  $\Sigma$  is assumed to be described by  state equations  in the form of (\ref{eq:Sigma}),   where  $x$,  $y$ and $z$ are respectively  the  plant state,  measured output and regulated output.  A known linear time-invariant exosystem $\Sigma_{exo}$ generates  the autonomous time-varying  reference signal $r$  and input disturbance signal $d$. The error  signal $e = z-r$ represents the difference between  the regulated output and the reference.    The exosystem can be written in the form of:
\begin{equation}
\label{eq:Sigma_exo}
\Sigma_{exo}:\;
 \left\{ \begin{array}{ccl}
\dot{w}(t) & = & S w(t) ,  \quad w(0) = w_0    \\
d(t)		& = & L_d w(t)  \\
r(t)		& = &   L_r w(t)
\end{array}\right.
\end{equation}
Here   $S$ represents the exosystem dynamics and $w$ is the state of the exosystem. Output matrices  $L_d$ and $L_r$ construct the disturbance and reference signals from the exosystem states.   By defining
\beann
E_w &=  &  H\,L_d  \\
D_{w} &=  & -L_r
\eeann  
we can replace  $\Sigma$ in  (\ref{eq:Sigma})  with the  error system $\Sigma_e$:
\bea
\Sigma_e: \left\{ \begin{array}{lcl}
	\dot{x}(t)   &  = & Ax(t)  +  Bu(t)  + E_w w(t),   \\
	\dot{w}(t)  &  =  & Sw(t),  \\
	  y(t)        &  =  &  C_y x(t)  \\
	  e(t)        &  =  & C_z x(t) + D_{w}\,w(t).
\end{array} \right. \label{Sigmae2}
\eea  
 A  feedback controller $u$ for the system $\Sigma_e$  is said to  achieve  \emph{exact output regulation} \cite{saberi2012control} if the closed-loop system is internally stable and,  for all initial  states $x_0$ and $w_0$ of the  plant and exosystem,  the system satisfies $\lim_{t \to \infty} e(t)= 0$.
Ensuring the error signal vanishes means that the input disturbance is  asymptotically rejected,  and the  controlled output
$z$  asymptotically tracks the desired reference signal $r$.
For the case where all  states  are measurable,  we have $y =x$ and state feedback can be used to  achieve  exact output regulation according to the following theorem:
\begin{theorem} \label{Thm31} \cite{saberi2012control}
	Assume system $\Sigma_e$ in (\ref{Sigmae2}) satisfies the following assumptions
	\begin{itemize}
		\item[(A.1)]  \label{A1} The pair $(A, B)$ is stabilizable.
		\item[(A.2)] The matrix $S$ is anti-Hurwitz-stable.
		\item[(A.3)] There \cyan{exist} matrices $\Gamma$ and $\Pi$ satisfying
		\begin{eqnarray}
		\Pi\,S & =  &  A\,\Pi+B\,\Gamma+E_w    \label{Pi1}\\
		0        &  =  & C_z\,\Pi +\red{D_{w}} \label{Pi2}.
		\end{eqnarray}
	\end{itemize}
	Let $F$ be any matrix such that $A + B\,F$ is  Hurwitz stable, and let $G = \Gamma-F\,\Pi$. Then the  state feedback control law
	\be
	\label{eq:FXGW}
	u  = Fx +Gw,
	\ee
	achieves exact output regulation for $\Sigma_e$.	
\end{theorem}
The Sylvester  matrix equations 	(\ref{Pi1})-(\ref{Pi2}) are known as the \textit{regulator equations} and generic solvability conditions are given in \cite{saberi2012control}.  The matrix $S$ is {\it  anti-Hurwitz stable}   if  none of its eigenvalues are stable. In fact, this assumption is not  essential and was   adopted in  \cite{saberi2012control} only to  avoid a   trivial problem  formulation in which output regulation is achieved by default because the exosystem  states vanish. 

In practice,  it is  not always  possible to measure all  states of the  plant,  and  an  estimate  $\hat x$ of the  plant state must be constructed  using  the  measured  output $y$.
%\begin{equation}
%\label{eq:ctrl_law}
%	\Sigma_c =  \left\{ \begin{array}{lcl}
%		\dot{v}(t) & = & A\,v(t)  +  B_c\,y(t)  \\
%		\dot{u}(t) & = & C_c\,v(t) +  D_c\,y(t)
%	\end{array} \right.
%\end{equation}
The following theorem gives conditions under which exact output regulation may be  achieved with  a dynamic measurement feedback controller.
%of the form  \ref{eq:ctrl_law}.
\begin{theorem} \label{Thm32} \cite{saberi2012control}
	Assume the system $\Sigma_e$ in (\ref{Sigmae2}) satisfies the assumptions (A.1)-(A.3). Further, assume
	the matrix  pair
	\[ \left( [C_y  \quad  0] ,	\left[
	\begin{array}{cc r}
	A     &   E_w     \\
	0     &      S
	\end{array}
	\right] \right) \]
	is detectable. Then the exact output regulation problem is solvable by a dynamic measurement feedback controller of the form
	\be
	\Sigma_c:
	\left\{ \hspace{-.2cm} \begin{array}{rll} %\bmat{ccc}
		\bmat{c} \dot{\hat{x}}(t) \\ \dot{w}(t) \emat &=& \bmat{cc} A & E_w \\ 0 & S \emat \bmat{c} \hat{x}(t) \\ w(t) \emat+\bmat{c} B \\ 0 \emat u(t) \\
		&+& \bmat{c} K_A \\ 0 \emat \Big( \bmat{cc} C_y & 0 \emat \bmat{c} \hat{x}(t) \\ w(t) \emat -   y(t) \Big) \\
		u(t) &=& F\hat{x}(t)+ Gw(t)
	\end{array} \right.
	\label{dolaw}
	\ee	
where $F$ and $K_A$  are such that $A+BF$ and  $ A + K_A C_y $ are both Hurwitz stable matrices,  and $G = \Gamma-F\,\Pi$.
\end{theorem}
In sections \ref{EOR_WT} and  \ref{sec:SimEnv},  we  discuss the  application of  EOR to  a  LIDAR-enhanced wind turbine.
For a given  mean wind speed $v_{\texttt{x},0}$,  we first develop a  state  model $\Sigma$    as in \eqref{eq:Sigma}, with  
coordinates  homogenised to the  mean  wind speed. 
The appropriate turbine model and controller  are used for Region  2  or  3 according to  whether the mean wind speed  is  below or above  the  rated wind speed of 11.4 $m/s$.
We then include wind dynamics  to  obtain the  error system $\Sigma_e$ as in \eqref{Sigmae2},  also in homogenised coordinates.

The vector $w$ represents the deviations  of the perpendicular wind speed (as measured by LIDAR)  from its  mean  value,  and its dynamics are  modelled by the matrix $S$.    The wind speed deviation is  modeled as an   input disturbance  $\bar d$ ,  and effective disturbance rejection means that the  effect of $\bar d$  on the turbine response is attenuated.   The  wind speed  deviation   also  determines the reference signal   $\bar r$. In Region  2,  the reference is the  rotor speed that  delivers the optimal  tip  speed ratio. In Region 3,  the reference is the blade  pitch angle that will maintain the  rotor speed at the turbine's rated power.

\red{Theorems \ref{Thm31} and \ref{Thm32} are  applicable to  the nominal linear  system $\Sigma_e$,  and hence the  application of the EOR control law  \eqref{dolaw} to  the nonlinear turbine  dynamics (as modelled by FAST)  cannot be expected to  achieve exact \cyan{output} regulation. 
However, the simulation results to  be presented  in Section \ref{sec:SimComp} will  show that the EOR control method achieves the required rotor speeds for  optimal energy  harvesting   with smoother control  actuation. This leads to reduced fatigue  loads, relative to alternative control methods that do  not incorporate  wind dynamics into  their controller design.} 

%\subsection{Disturbance Accomodarion Control (DAC)}

\subsection{Alternative Turbine  Control Methodologies} \label{secAltCont}
To  demonstrate the effectiveness  of  the EOR method in  reducing fatigue loads, in 
Section \ref{sec:SimComp} we shall  compare  its  performance  with  two alternative control methods  whose application to  turbines have been widely studied. Here, we briefly describe these  alternative methods.

One of the most well-known wind turbine control  methods is  DAC,  which uses wind estimation data  to reduce  or cancel  persistent input disturbances. Similar to EOR, DAC assumes a linear  plant  model of the form (\ref{eq:Sigma}).  Disturbance states are created  by the augmentation of a state-estimator in a state feedback controller by an assumed-waveform model. These disturbance states are used to reduce or counteract the persistent disturbance effects of the wind.  The wave-form model is commonly assumed to  take  a  constant value (\cite{wright2004modern, wright2008advanced,balas2011adaptive}) in the  form
\begin{eqnarray}
	\dot{z}_d(t) & = & 0, \label{DACzeq} \\
	d(t) & = & z_d(t),   \label{DACdeq}
\end{eqnarray}
where $z_d(t)$ is the state of the disturbance model.
In  this paper,  the LIDAR-enhanced variant of DAC   known  as DAC+LIDAR  \cite{wang2013comparison}
will be used for comparisons, in  which $d(t)$  is taken as the LIDAR longitudinal wind measurement. 
%In Section IV we describe the application of  EOR to  a turbine using  LIDAR measurements.
 Thus in our performance comparisons in Section VI, EOR and DAC+LIDAR have access to the same wind preview information. 
 \red{For simplicity, in the following we shall  use DAC when \cyan{referring} to the  LIDAR-enhanced version}. 
The DAC control law is
\begin{equation}
\label{eq:DAC}
u(t) = Fx(t)+G_d z_d(t) ,
\end{equation}
where  $F$ is a  state feedback matrix chosen to  place the closed-loop poles at certain desired locations,  and $x$ is the state variable of the wind turbine model. Then, if possible, the wind disturbance state gain $G_d$ is  chosen to exactly cancel out the wind disturbance  by solving
\begin{equation}
\label{eq:DACSolve}
B G_d + H   = 0
\end{equation}
for $G_d$ in which $B$ and $H$ are defined in  (\ref{eq:Sigma}).

\red{When \eqref{eq:DACSolve} is solvable,  the DAC control methodology may be viewed as a special case of the EOR control methodology.} If we apply the  simplifying assumptions  $S=0$   and  $L_d =1$   to  (\ref{eq:Sigma_exo}), we obtain $E_w = H$  and
\begin{eqnarray}
\dot{w}(t) & = & 0 \label{DACeq1} \\
d(t)  &  = & z_d(t)  \label{DACeq2}
\end{eqnarray}
which are identical to (\ref{DACzeq})-(\ref{DACdeq})  with  $w = z_d$.  Using   $\Pi =0$ in the EOR control law gives   $\Gamma = G_d$, so   the first  regulator equation (\ref{Pi1}) becomes
\be
0 =  B G_d + H  \label{DACeq3}
\ee
which is  (\ref{eq:DACSolve}).

The equation   (\ref{eq:DACSolve}) may not be solvable if the vector $B$ has zero elements.
Then   $G_d$ is chosen by the  minimization problem
\begin{equation}
\label{eq:DAC_min}	
\mbox{argmin}_{G_d} \quad \| B G_d + H \|_2.
\end{equation}
In such cases, DAC will not be able to exactly cancel the disturbances.

We note a number  of similarities and differences between EOR and DAC. Both use a  state feedback law   to  stabilise the closed-loop dynamics and a feedforward term to  cancel  input disturbances.  Where DAC models  the disturbance as a constant,  the dynamic exosystem used in  EOR enables  greater flexibility in the modelling of the disturbance. Moreover,  under mild system assumptions  of controllability and observability,   \eqref{Pi1}-\eqref{Pi2}  are  solvable. By  contrast,  only    approximate solutions can  be  obtained for  (\ref{eq:DACSolve}), leading to only approximate disturbance cancellation.  Additionally, DAC cannot  ensure the output   tracks any desired time-varying reference.

Hence we can expect  better disturbance rejection performance from EOR as it has access to  derivatives of the disturbance input. Conversely,  DAC only has access to the absolute value of the disturbance. Moreover,  non-solvability of the DAC minimization equation \eqref{eq:DAC_min}    can occur   if  the disturbance input vector $H$ and control input vector $B$ are  orthogonal.  In such  cases, obtaining  a non-zero solution for $G_d$ will  require some plant model reduction,  leading to reduced control performance.

In our simulation   results,  we shall  also  compare the  control performance of EOR and DAC against
a Baseline method of wind turbine control, employing  a proportional torque controller  of the  kind  commonly  used in \cyan{the industry}.  The standard (Baseline) controller for Region 2 is a generator torque reference proportional to the square of the rotational speed of the rotor
\begin{equation}
\label{eq:BL_R2}
M_g = k \Omega_r ^ 2,
\end{equation}
where $k$ is given by
\begin{equation}
k = \frac{1}{2} \rho \pi R^5 \frac{C_{p}(\lambda^*,0)}{(\lambda^*)^3}.
\end{equation}
%and $\lambda_*$ as mentioned, is the optimal tip speed ratio which imposes $C_p = C_{p_{max}}$ \cite{pao2011control}. \\
In Region 3, the  Baseline controller is a PI regulator which is set to eliminate rotor speed error by generating the required references for the blade pitch angle. The feedback information is taken from the rotor speed $\Omega_r$ and compared against the rated rotor speed. The  rotor speed error is then fed into a conventional PI controller to generate the pitch command $\theta_c$ in the following form
\begin{eqnarray}  %fullmodel
\label{eq:PID}	
\Delta \Omega &=  &\Omega_r - \Omega_0\\
\theta_c &= & K_p \Delta \Omega + K_i \int \Delta \Omega dt
\end{eqnarray}
The design procedure  for determining the proportional and integral gains $K_p$ and $K_i$ is described in \cite{wright2008advanced}.

%%%%%%%%%%%%%%%%%%%%%%%%%%%%%%%%%%%%%%%%%%%%%%%%%%%%%%%%%%%%%%%%%%%%%%%%%%%%%%%%%%%%%%%%%%%%%%%%%%%%%%%%%%%%%%%%%%%%%%%%%%%%%%%%%%%%%%%

\section{Synthesizing Disturbance and Reference Exosystems for Wind Turbines}
\label{EOR_WT}
A key component of the EOR control  methodology  introduced in Section \ref{secEOR} is a linear exosystem  in \eqref{eq:Sigma_exo} that generates  the known disturbance and reference  signals.    In this section, we describe how LIDAR measurement data  can be used to synthesize linear exosystem dynamics to represent  short-term wind evolution with a high fidelity.
%
%  The availability of wind  preview information  enables the  wind to be  modelled as a known  exosystem generating input disturbances and reference signals.  Disturbance rejection means that the effect of wind variations on the desired rotor speed is rejected,  and successful  reference tracking means that  the turbine  archives the optimal tip speed ratio to ensure the maximum power generation.
% Since LIDAR can provide look ahead information of the wind, the controller has enough time to construct these approximated subsystems and solve the equation \eqref{Pi1} and \eqref{Pi2}. By that, it can be expected that the turbine actuators such as Generator torque and blade pitch angle, can be performed without their associated time constant having a negative effect in the turbine's reaction to the variation of the wind.
%Disturbance and reference generating exo-system $\Sigma_{exo}$ represents the LIDAR-measured upcoming disturbance and  the reference signal for the generator torque $\tau_c$ and/or for the pitch angle depending the region of operation. They can be put in the form of a single exo-system  \eqref{eq:Sigma_exo} with two different output outputs represented by output gain vector $L_d$ and $L_r$. % Next, we  obtain feedback matrices $F$ and $K_A$ for the  measurement  feedback  controller  $\Sigma_c$, plant and exo-system observers.

If $f$  is the focal distance of the CW-LIDAR and $v_{\texttt{x},0}$ is the mean longitudinal wind speed, a time window of $T_f = \frac{f}{v_{\texttt{x},0}}$  of wind preview  information  will be  available.   Since $T_f$ depends on the mean wind speed, a constant preview length of $0 < T_{pl} < T_f$ is assumed to cover the whole range. In this work, we assume  $f = 60\ m$ and  our largest wind speed  is  $24\  m/s$, hence  $T_{pl} = 1.5 \ s$ is suitable.  To find an exosystem that can accurately model the longitudinal wind signal $v_{\texttt x}$, an auto-regressive model is fitted to the longitudinal wind speed signal $\hat v_{\texttt x}$ provided by the LIDAR (our LIDAR model will be discussed in Section \ref{subsec:CWLIDAR}) over the time  window $[0, T_{pl} ]$. Therefore, the combined exosystem and  turbine  model
\eqref{eq:Sigma} are constructed in discrete time. The exosystem dynamics are modeled with an LTI difference equation
%The result is a difference equation which may be written as the following polynomial
%	\vspace{-1mm}
			\begin{equation}
		\label{eq:ARMA}
			w_{\texttt x}[n] = a_1 w_{\texttt x}[n-1] + a_2 w_{\texttt x}[n-2] + \dots + a_{{\red N}} w_{\texttt x}[n-{\red N}],
		\end{equation}
		in which
		\be
		w_{\texttt x}[n] = \hat{v}_\texttt{x}[n] - \hat v_{\texttt{x},0}  \label{wxsignal}
		\ee is the deviation of the LIDAR-measured longitudinal  wind speed  $\hat{v}_x$ from the  mean LIDAR longitudinal wind speed  $ \hat v_{\texttt{x},0}$,  at  sampling instant  $n$. The $a_k$ are the system coefficients and $N$ is the order of exosystem dynamics to be chosen. 

%Assigning  $N$ requires knowing two factors of desired tracking quality and also the plant dynamics.
Choosing higher values of $N$ yields better representation of the wind deviation, leading to better disturbance rejection and reference tracking.
%When disturbances and the control inputs are unmatched, (i.e, when $d$ enters from a channel where there is no direct access with $u$),  higher derivatives of  $d$ are necessary for effective disturbance rejection.
However, aggressive disturbance rejection and reference tracking are contradictory to fatigue load reduction.  This is because perfect disturbance rejection would require all the turbine states to precisely track their equilibrium value against the wind speed, which would increase  tower deflection and  drive-train torsion.
% Higher-order difference equations also increase the  computational burden.
%Therefore, a lower bound for $N$ depending the plant dynamics and state space equations exists which regarding the used wind turbine plant model in this work is $N \ge 2$.
Thus, a trade-off needs to be made between higher fidelity modeling of the wind (using larger $N$ values) and reducing the time-variation of the control input signal (using lower $N$ values).

Throughout this work, $N={\red 2}$ has been empirically chosen to generate exosystems that are computable in real time. Hence we  introduce a  state vector for  the exosystem  as
  $w[n]=  [ w_{\texttt x}[n], w_{\texttt x}[n-1],  w_{\texttt x}[n-2] ]^T$,  and  the exosystem  dynamics are given  by
\begin{eqnarray}
	w[n+1] \hspace{-3.5mm} & = &  \hspace{-3.5mm}
	\underbrace{
		\begin{bmatrix}
			a_{1}^* &a_{2}^* & a_{3}^* \\
			1        &  0        & 0                \\
			0        & 1        & 0
	\end{bmatrix}}_{S} \hspace{-1mm} 	w[n],
	\label{eq:exoHomogen}
\end{eqnarray} 

\begin{figure}[t]
\centering
	\includegraphics[width=10cm]{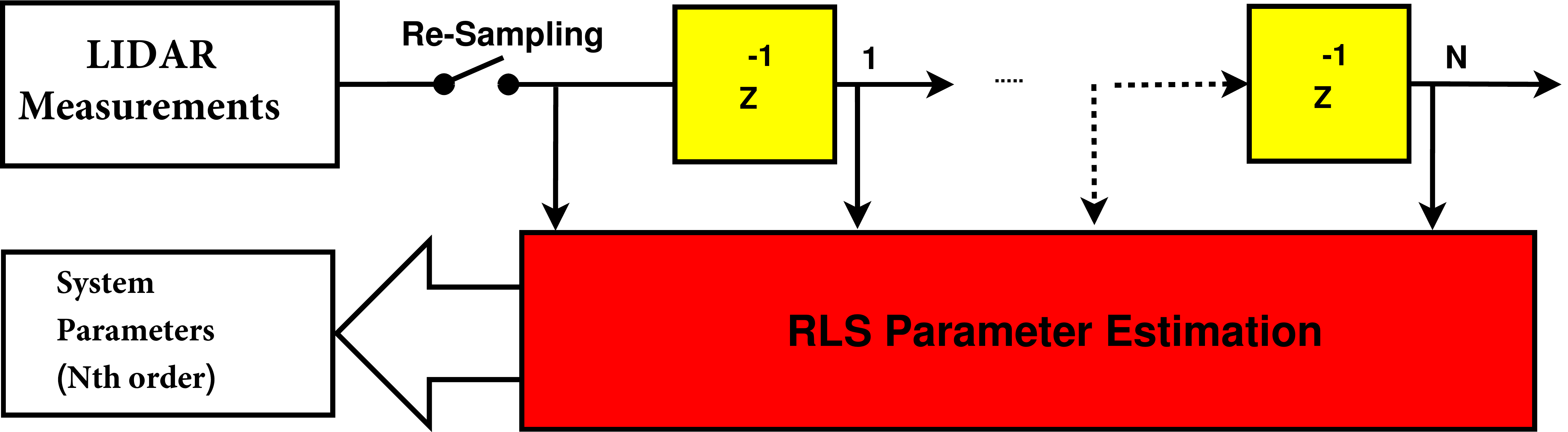}
	\caption{Recursive least square estimation of parameters}	\label{fig:RLS}
\end{figure}

The  optimal coefficients $a_{k}^*$,  for $1 \leq k \leq  3$ \red{ may be obtained by  passing a batch of sampled  LIDAR measurements  into a Least Square Algorithm  every $T_{pl}$ seconds.  However, a batch least square method has to perform large matrix inversions at every $G$ matrix update event  which might be not feasible for real-time applications. 
By using Recursive Least Square (RLS) instead, this computation load is distributed evenly for each simulation step. The RLS estimator for $a_{k}^*$ coefficients are shown in Figure \ref{fig:RLS}. Thus the $a_k$ coefficinets, and hence the  $S$ matrix, are updated at   each time  step,  with  frequency 10 Hz. With a forgetting factor of 0.90 in the RLS, the weight of each sampled data becomes almost zero after 6 seconds, as $0.90^{60} \approx 0.001$. 
The regulator equations \eqref{Pi1}-\eqref{Pi2}, and consequently the feed-forward gain matrix  $G$, are updated every 6 seconds.}

\begin{figure}[t]
\centering
	\includegraphics[width=.6\textwidth]{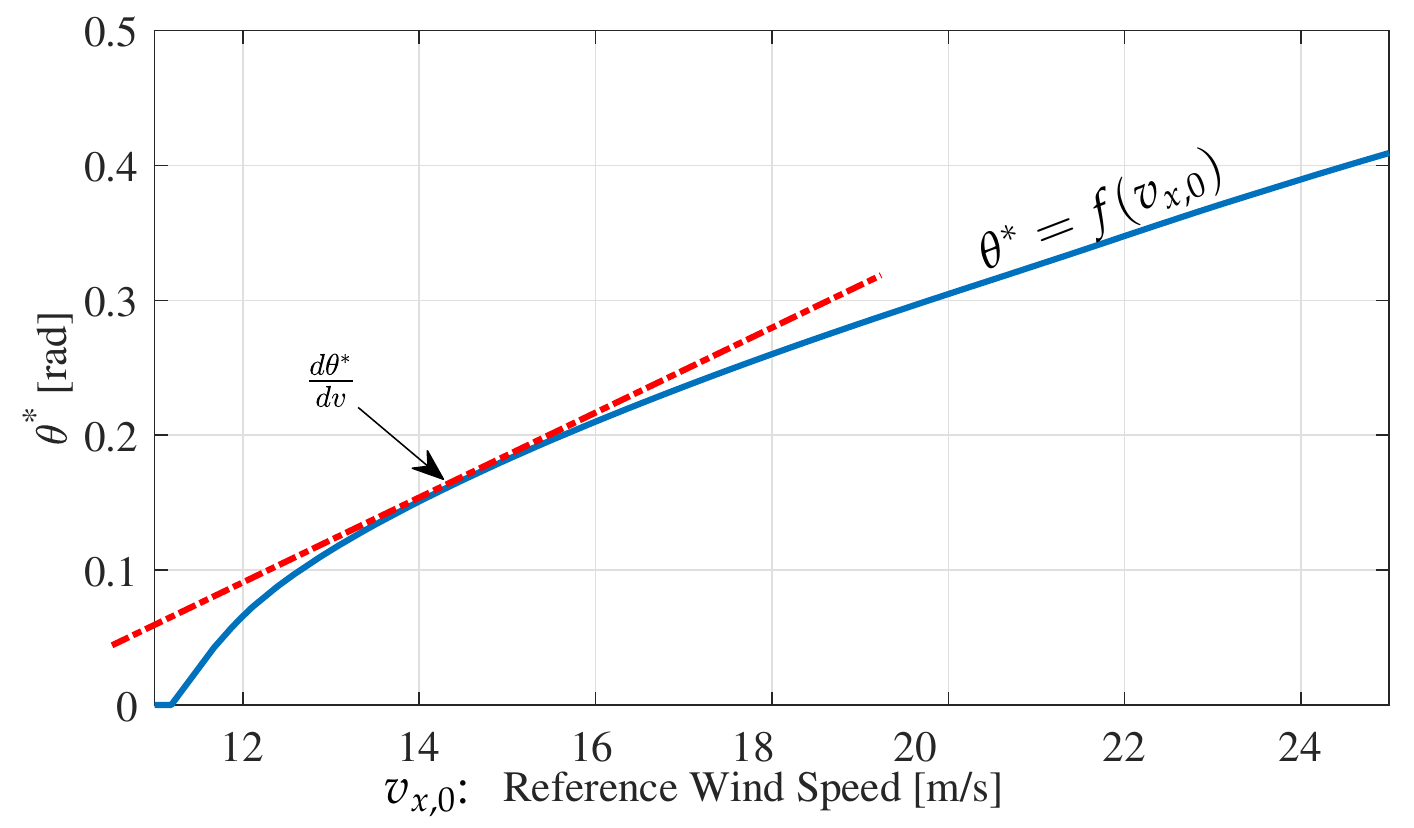}
	\caption{Equilibrium locus of the optimal  blade  pitch angle  $\theta^*$  for Region 3 wind speeds (Blue) and  its tangent (Red)}  	\label{fig:ssv_Graph}
	\end{figure}

To  complete the exosystem model  $\Sigma_{exo}$ in  \eqref{eq:Sigma_exo}, output matrices $L_d$ and $L_r$ should be also determined.
 The disturbance input   is  the deviation of the  longitudinal  wind speed  from  the  mean value
 $\bar{d}(t) = \hat{v}_{\texttt{x}}(t) -\hat{ v}_{\texttt{x},0}(t) = w_{\texttt x}(t)$,  so
\be
L_d  = [1 \quad 0 \quad 0].
\ee
As discussed in Section \ref{SecOPG}, in Region  2  the controlled output is  the  rotor  speed $\Omega_r^*$,  and from \eqref{eq:tsr}  we  have, in  homogenised coordinates,
$\bar{\Omega}_r = \frac{\lambda^* w_{\texttt x}}{R}$.  Taking $\bar \Omega_r(t)$ as the time-varying reference signal  to  be tracked by the EOR controller, we need
\be  L_r =  \left[\frac{\lambda^*}{R}  \quad 0 \quad 0 \right]
\ee
		%At each linearization point as shown in Figure \ref{fig:ssv_Graph}, a tangent line locally describes the variation of the desired state regarding to the wind. 	
	For a Region 3  mean longitudinal wind speed $v_{\texttt{x},0}$,  the controlled output is the  blade  pitch angle  $\theta^*$, obtained by solving \eqref{theta*}. 
Figure \ref{fig:ssv_Graph} shows the graph  of  $\theta^* $ as a function of $v_{\texttt{x},0}$.  

 To  obtain the time-varying  reference signal  $\bar \theta(t)$ in  homogenised coordinates,   we use the first  order approximation
	\be
	\bar \theta  = \frac{d \theta^*}{d v_{\texttt{x},0}} w_{\texttt x}
	 \ee
 Thus  for  Region 3,
\be L_r =  \left[ \frac{d \theta^*}{d v_{\texttt{x},0}} \quad 0 \quad 0 \right] \ee
%Combining $S$, $L_r$ and $L_d$ yields the exosystem $\Sigma_{exo}$ in \eqref{eq:Sigma_exo} that will be used to develop the EOR dynamic measurement feedback controller $\Sigma_c$ for the turbine.

%\subsection{Plant and Exo-System State Estimator}

%%%%%%%%%%%%%%%%%%%%%%%%%%%%%%%%%%%%%%%%%%%%%%%%%%%%%%%%%%%%%%%%%%%%%%%%%%%%%%%%%%%%%%%%%%%%%%%%%%%%%%%%%%%%%%%%%%%%%%%%%%%%%%%%%%%%%%%%%%%%

\section{Turbine Simulation Environment}
\label{sec:SimEnv}

\begin{figure}[t]	
\centering
	\includegraphics[width=.6\textwidth]{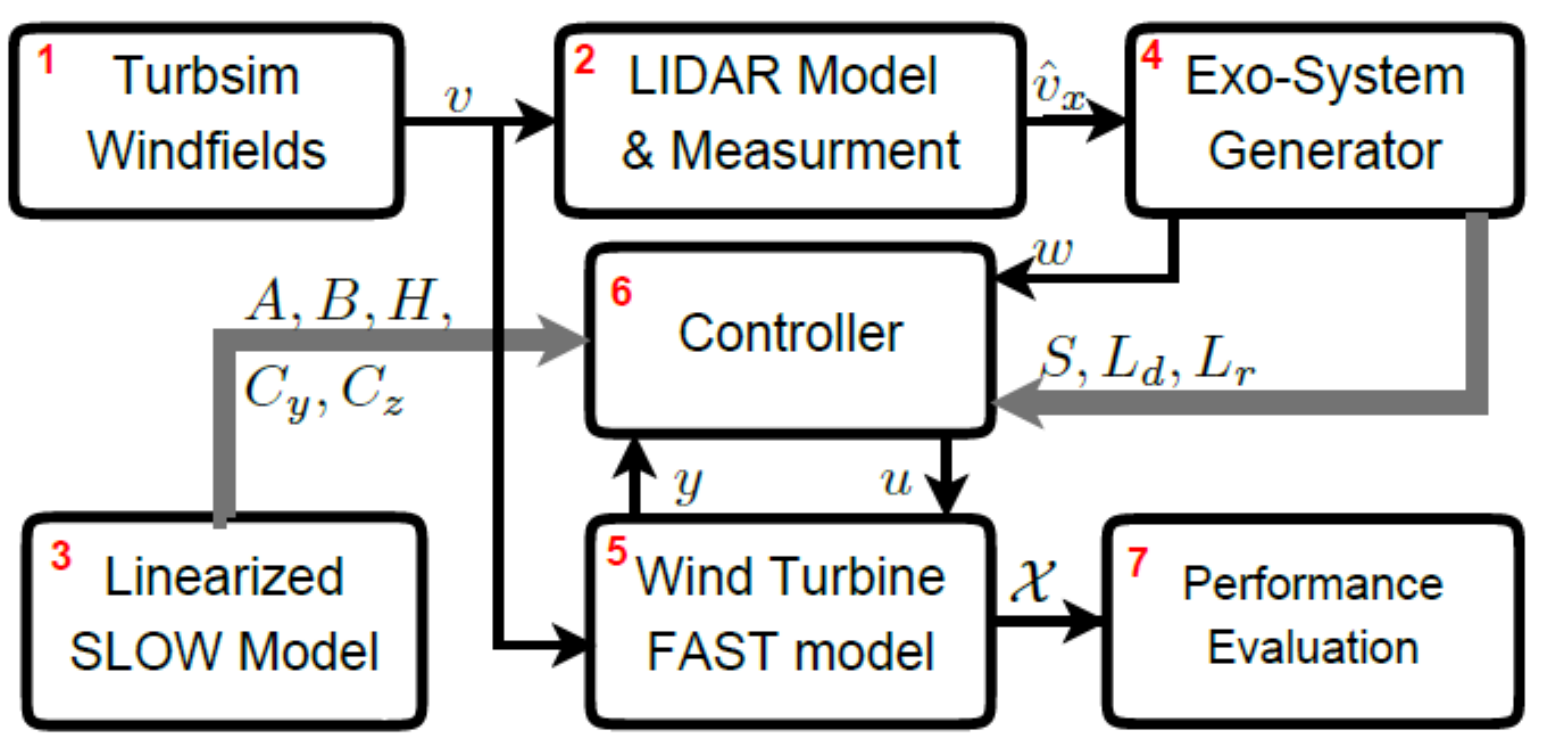}
\caption{Block Diagram of TOR. Thick gray lines represent matrices and thin black lines represent signals.}  \label{TOR}
\end{figure}

In this section, we describe our   turbine simulation environment,  developed in $\text{Simulink}^{\textregistered}$,
and  named the \textit{Turbine Output Regulator} (TOR). Its purpose is the   simulation and comparison of the control performance of  the  EOR, DAC and  Baseline  control methodologies for the NREL 5 $MW$ FAST turbine model. The block diagram of TOR is illustrated in Figure \ref{TOR}. It is comprised of seven subsystems as follows:
\begin{enumerate}
\item  The TurbSim package \cite{jonkman2009TurbSim} for the simulation of  realistic wind fields.
\item The LIDAR simulator based on  \cite{dunne2011lidar}.
\item  The  linearized model of the nonlinear Simplified Low-Order of Wind (SLOW) Turbine model.\
\item The linear exosystem generator obtained from LIDAR data, as described in Section \ref{EOR_WT}.
\item The high fidelity wind turbine simulator FAST.
\item  The controller subsystem implements  the Baseline,   DAC and EOR controllers described in Section \ref{sec:WCon}.
\item  Performance measurement code to   calculate  metrics  related to  the  power generation  and DELs.
\end{enumerate}
%and the solver for  regulator equations of \ref{Pi1} and \ref{Pi2} to obtain the feedforward gain vector $G$ in \eqref{eq:FXGW}. The state feedback vector $F$ in \eqref{eq:FXGW} can be obtained by LQR algorithm  or any other arbitrary pole placement method. Determining the observer gain $K_A$ and estimating the plant states are carried out by a Kalman filter.

We  now briefly describe  each subsystem:

\subsubsection{TurbSim Wind Field  Simulator} \label{secTurbSim}

\begin{figure}[t]
\centering
	\includegraphics[trim={16mm 0 0 2mm},clip, height=5cm ]{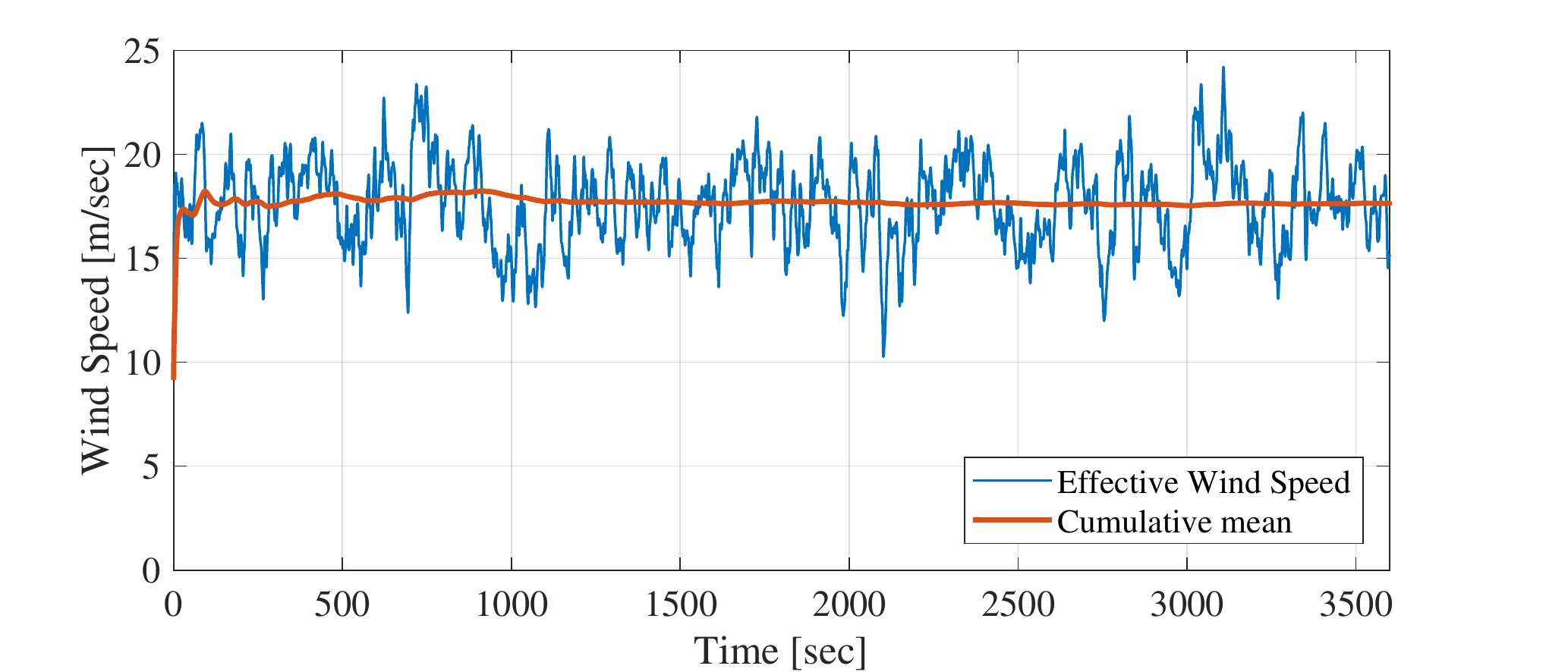}
	\put(-270,40){\rotatebox{90}{\scriptsize	 Wind speed (m/s)}}	
	\caption{Effective Hub-height  wind speed of Class-A TurbSim  wind signal  and  its  cumulative mean.  }
	\label{fig:Rewf_Wind}
\end{figure}

TurbSim  is a full-field, turbulent-wind simulator developed by NREL  using stochastic models to generate realistic three-dimensional wind field vectors $v$,   with components for  the longitudinal, crosswise and vertical components of the wind,  in arbitrary resolution.  The detailed parameters of wind input files are the vertical stability parameter $Ri_{TL}$, shear exponents $\alpha_D$ and the  mean friction velocity $u^\star_D$ \cite{kelley1999case} which are set on the default values in TurbSim to generate 1 hour of wind information.  Figure \ref{fig:Rewf_Wind} shows  a TurbSim-generated Class-A intensity  wind signal  of one hours duration,  with a mean longitudinal wind speed of $v_{\texttt{x},0} =18 \ m/s$. We also  show the  cumulative  mean  of the   signal. The output $v$ from TurbSim  in  Figure \ref{TOR} is the wind  field vector which is  applied to both the FAST turbine simulator  and the LIDAR simulator.

The 5 $MW$ reference wind turbine is a class-A wind turbine, and consequently,  the  FAST simulator  is exposed to a broad range of Class-A intensity wind fields generated by TurbSim  according to IEC-61400-1 standard \cite{IEC}.  These wind fields have mean longitudinal wind speeds from 8 to 24 $m/s$ with resolution steps of 2 $m/s$. Wind speeds below  8 $m/s$ are not  considered,  as \cite{jonkman2009definition} does not  recommend the  use of  a Baseline torque controller of the form \eqref{eq:BL_R2} below this wind speed.

\subsubsection{LIDAR Simulator}
\label{subsec:CWLIDAR}
\begin{figure}[t]
	\centering
	\includegraphics[width=7cm]{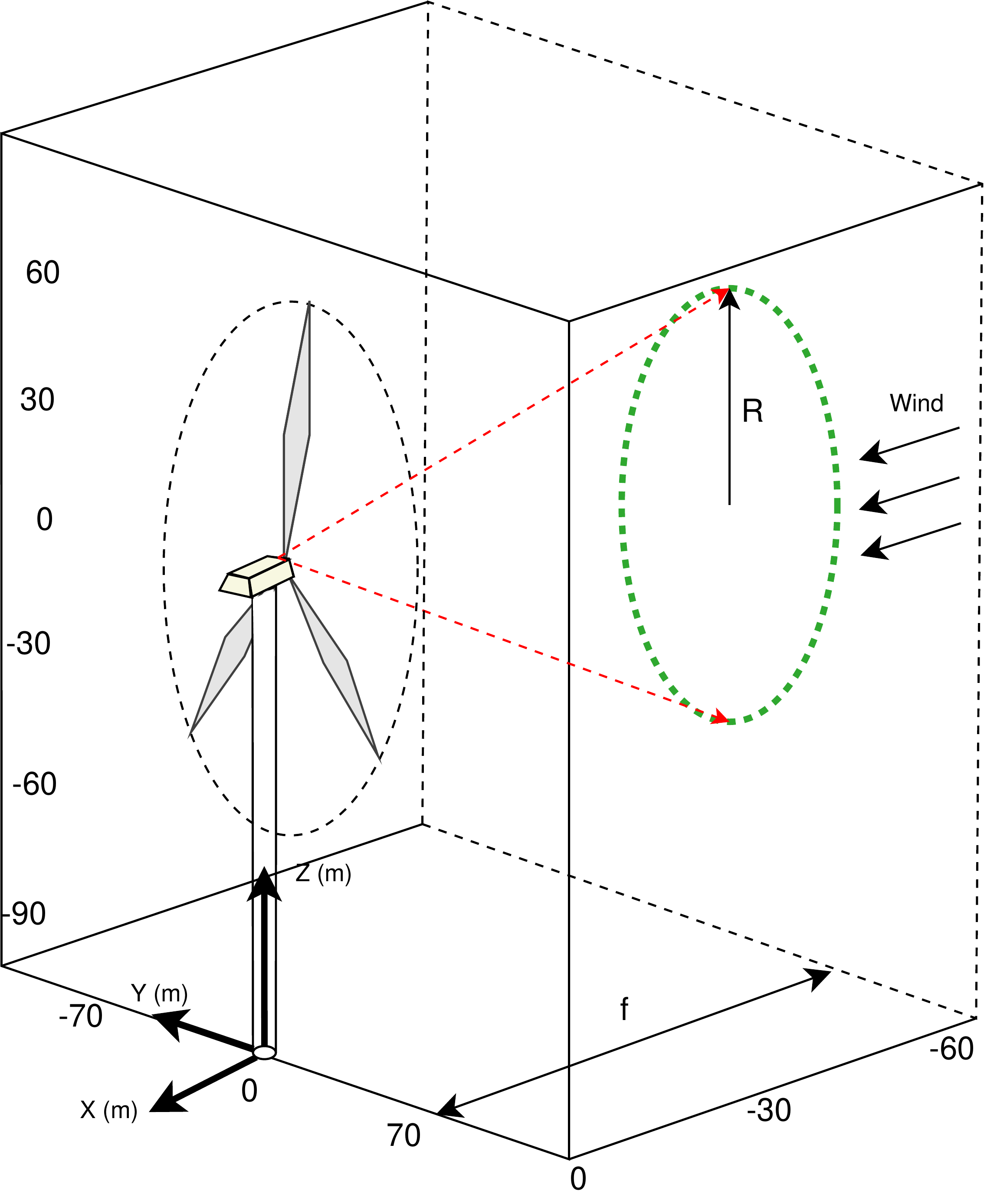}
	\caption{Coordinate system of wind turbine and nacelle mounted LIDAR shown by $\texttt x,\texttt y$ and $\texttt z$. The distance of the focal point of CW-LIDAR is shown by $f$ and $R$ is the scan radius. }
	\label{fig:LIDAR}
\end{figure}

We use the  continuous wave CW-LIDAR model described in \cite{dunne2011lidar} to simulate the  longitudinal wind speed $v_{\texttt x}$  at a specific distance from the turbine  blades by focusing the laser beam at that location.
Figure  \ref{fig:LIDAR} depicts the coordinate system and geometrics of the LIDAR placement on the wind turbine nacelle.

\begin{figure}[t]
	\centering
	\includegraphics[trim={3.4cm 0 0 10mm},clip,height=5cm]{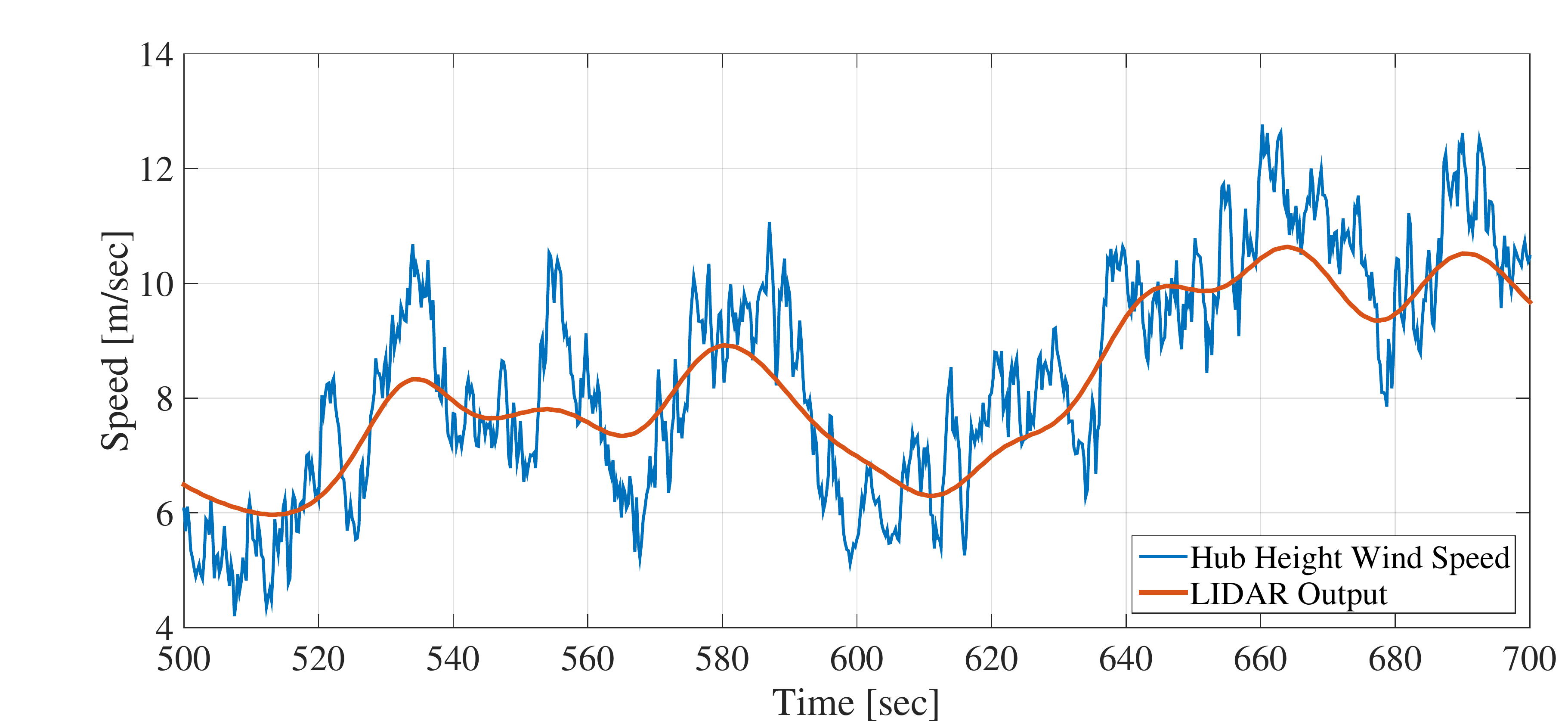}
	\put(-257,40){\rotatebox{90}{\scriptsize	 Wind speed (m/s)}}	
	\caption{Blue: A 200 Seconds sample of a Region 2 Class-A wind with mean speed of 8 $m/s$, measured at the hub height. Red: Output of the CW-LIDAR model.} % EKF as estimated states of the disturbance exo-system.
	\label{fig:LIDAR_LPF}
\end{figure}

Twenty-four evenly distributed  measuring points on  a circular cross-section of the wind vector  $v$  at focal distance $f$ from  the  rotor plane are scanned by the  LIDAR beam,  and spatial  averaging is applied to  the wind speeds along the  length  of each measurement beam \cite{dunne2011lidar}. The  effect of the  averaging is equivalent to passing the wind  signal  through  a  non-phase distorting low-pass filter whose $3\ dB$ bandwidth is determined by
\begin{equation}
\label{eq:LIDAR_BW}
BW_{3dB} = \frac{87}{f^2}.
\end{equation}
The constant 87 is based on specific parameters of the LIDAR used in   \cite{dunne2011lidar}.
Averaging Riemannian sums of the  24 measurements in the cross-section yields  $\hat v_{\texttt x}$,  an approximation to the  longitudinal speed  $v_{\texttt x}$. In Figure \ref{fig:LIDAR_LPF}  the overlay comparison between the real hub-height wind signal and the simulated  LIDAR output $\hat v_{\texttt x}$ with a focal distance of 60 metres is shown.

 TurbSim generates  wind signals according to Taylor's Frozen Wind
Hypothesis,  which models the wind field as a turbulence box moving  towards the wind turbine at  its average wind speed.
 Thus  the wind field is assumed not to  evolve between the LIDAR focal point and  the blades. This  hypothesis is appropriate  for relatively flat terrain where  geological features do not interact  with the air flow between the measurement point and blades.  \red{Additionally,  the induction zone caused by the turbine itself certainly causes some  wind evolution between the LIDAR focal point and the blades}. 

\subsubsection{Linearized low order model}

In order to  compute a low-order linearized turbine model, knowledge of the  mean  wind speed  $ v_{\texttt{x},0}$ is required.   It is apparent from Figure \ref{fig:Rewf_Wind} that after some 100  seconds,  the cumulative mean of the longitudinal  wind signal $v_{\texttt x}$ gives a good approximation to the mean longitudinal wind speed $v_{\texttt{x},0}$,  and hence we use   the cumulative mean,  denoted by $\hat v_{\texttt{x},0}$, as the linearization point satisfying \eqref{eq:eql_point}.   The linearized model \eqref{eq:Sigma} is  obtained as  described in Section \ref{ssec:LWTm},  based on the parameters given in Table \ref{table:5MW}  in  coordinates homogenised to  the  assumed mean wind speed.   This  linear model is passed to  the controller subsystem for  the computation of the DAC and EOR control inputs.

\subsubsection{Wind Exosystem Generator}
This subsystem uses the cumulative mean wind speed to   construct the   homogenised   wind signal $w_{\texttt x}$ in
\eqref{wxsignal}. Matrices $S$, $L_r$ and $L_d$ for the  exosystem $\Sigma_{exo}$  \eqref{eq:Sigma_exo}, are obtained as   in Section \ref{EOR_WT}.    The exosystem  state variable  $w$,  disturbance  $d$ and  reference $r$ are  passed  to the controller subsystem.

\subsubsection{{FAST Turbine Simulator}}

To  simulate the response of the 5 $MW$  HAWT turbine, we use the NREL FAST 7 code.  A compiled MATLAB$^{\textregistered}$ S-Function of the FAST code is used to link the designed controllers to  FAST in Simulink$^{\textregistered}$.  Table \ref{table:DOFs}  lists the
  degrees of  freedom  applicable  to the on-shore 5 $MW$ wind turbine that have been activated in FAST for our simulation.  FAST does not include a model for pitch actuator, so a second order servo-system according to Table \ref{table:5MW} has been added to the  Simulink$^{\textregistered}$ environment.

The FAST model  receives the  wind  field vector $v$ from  TurbSim  and  uses  it to  compute the turbine state vector  $\bar x$  and  measured  output signal  $\bar y$. These signals are passed  to  $\Sigma_c$ in \eqref{dolaw} for computation of the control input  $\bar u$.
Finally, all  measurements contained in the FAST output log files, represented by ${\cal X}$ in  Figure  \ref{TOR},  are  passed to  the  performance measurement  subsystem. These include  the blade root and tower root bending moments, rotor and generator speeds, and torsion of the drive shaft,  as well  as the pitch angle  and  generator torque input commands.
\begin{table}
	\caption{Enabled DOFs in FAST code}
	\label{table:DOFs}
	\centering
	\begin{tabular}{p{4cm} c r}
		\hline\hline
		Enabled mode                                 & No. of DOFs  & Total \\ \hline
		%%%
		Generator                              & 1            & 1     \\
		Drive-train Torsion                          & 1            & 1     \\
		1$^{st}$ \& 2$^{nd}$ fore-aft tower bending  & 1            & 2     \\
		1$^{st}$ \& 2$^{nd}$ side-side tower bending & 1            & 2     \\
		1$^{st}$ edge-wise blade                     & 1 $\times$ 3 & 3     \\
		2$^{nd}$ edge-wise blade                     & 1 $\times$ 3 & 3     \\
		1$^{st}$ flap-wise blade                     & 1 $\times$ 3 & 3     \\
		2$^{nd}$ flap-wise blade                     & 1 $\times$ 3 & 3     \\ \hline
		Total DOFs                                   &              & 18    \\ \hline
	\end{tabular}
\end{table}	

\subsubsection{Controller subsystem}  This subsystem  computes the EOR, DAC and Baseline controller input signals $u$,  and  passes them to  the  FAST simulator.
 The EOR dynamic measurement feedback controller  $\Sigma_c$ is given by \eqref{dolaw}. The state feedback matrix $F$ is chosen by the  LQR algorithm  with $Q=C_z^T C_z$,
% where  $C_z$ is given  by either \eqref{CzReg2} or \eqref{CzReg3}, as appropriate for the current  operating region,
  and $R$ is empirically chosen to avoid control input saturation.  The same matrix $F$ is used for  the DAC controller.  The  estimated  homogenised state vector $\hat {\bar x}$ is obtained  using the homogenised measured outputs $\bar y$ received from  the FAST simulator.   The  observer  feedback gain  matrix $K_A$ is determined by a Kalman filter. 
  The wind turbine model in \eqref{eq:Sigma} does not have a process noise term, and  the output measurements obtained from the outputs of the FAST contain only numerical errors,  so   the process noise covariance matrix of the Kalman filter can be neglected,  and  a very small  measurement noise covariance matrix  is sufficient.
\subsubsection{Performance Measurement}
 The  damage equivalent loads computed are the  tower root fore-aft bending moment $M_{\texttt{y}T}$,  tower root side-to-side bending  moment $M_{\texttt{x}T}$,
 blade root flapwise bending moment $M_{\texttt{y}B}$,  blade root edgewise bending  moment $M_{\texttt{x}B}$, and $LSS$ torque.  These are computed using  the Rain-Flow-Counting method \cite{downing1982simple}, with DEL computations  performed with the Rain-Flow-Counting-Algorithm open source $\text{MATLAB}^{\textregistered}$  code \cite{RFCToolbox}.
This subsystem also  computes the average power generated  $P_{mean}$,  and for  Region  3  operation  we compute the power standard deviation  $std(P)$ and  rotor speed standard deviation $std(\Omega_r)$. Smaller power standard deviation indicates the  power generation is maintained  close to the rated value of 5  $MW $.  For Region  2  operation we compute the tip speed ratio standard deviation $std(\lambda)$, with smaller values  indicating  better tracking of  the optimal tip speed ratio $\lambda^*$.

Additionally,  the measurement subsystem provides spectral analysis of  the tower fore-aft bending moment,  tower side-to-side bending moment, blade flap-wise bending moment, low speed shaft torque, pitch rate and the generated power.  Reduced high frequency content in  the  power spectral density (PSD) of these variables implies reduced fluctuations of the measured variable.  For bending moment (tower or blade) signals,  the integral of the amplitude of the PSD over the frequency range is an indicator of the wind energy dissipated within the turbine tower or blade.   The consequence of this energy dissipation is fatigue accumulation in the tower or blade,  and thus reduced
PSD amplitudes are associated with lower lifetime turbine damage.

The TOR environment is able  to  compute  one  hour  of turbine response simulation in  20 minutes of CPU time on  a contemporary desktop PC, and
%Moreover, this  CPU time includes   the computation time of the FAST code for the  turbine response,  in  addition  to the computation of the EOR control input.
 hence  the output regulation control methodology can be expected to be suitable for real-time realization on a wind turbine.

%%%%%%%%%%%%%%%%%%%%%%%%%%%%%%%%%%%%%%%%%%%%%%%%%%%%%%%%%%%%%%%%%%%%%%%%%%%%%%%%%%%%%%%%%%%%%%%%%%%%%%%%%%%%%%%%%%%%%%%%%%%%%%%%%%%%%%%

\section{Results and Comparisons}
\label{sec:SimComp}
We  now present the  simulation results from our investigation of the performance of the  EOR, DAC and Baseline controllers introduced in Section \ref{sec:WCon}.   The simulations use class-A turbulent wind signals with  mean wind speeds ranging  between 8 $m/s$ and 24 $m/s$,  as described in Section \ref{secTurbSim}. The first 100  seconds  of turbine  response are excluded from  the comparisons,  as this time  is  required for the  initialization of the  DAC and  EOR controllers.

	\begin{figure}[t]
		\centering
		\includegraphics[width=9cm]{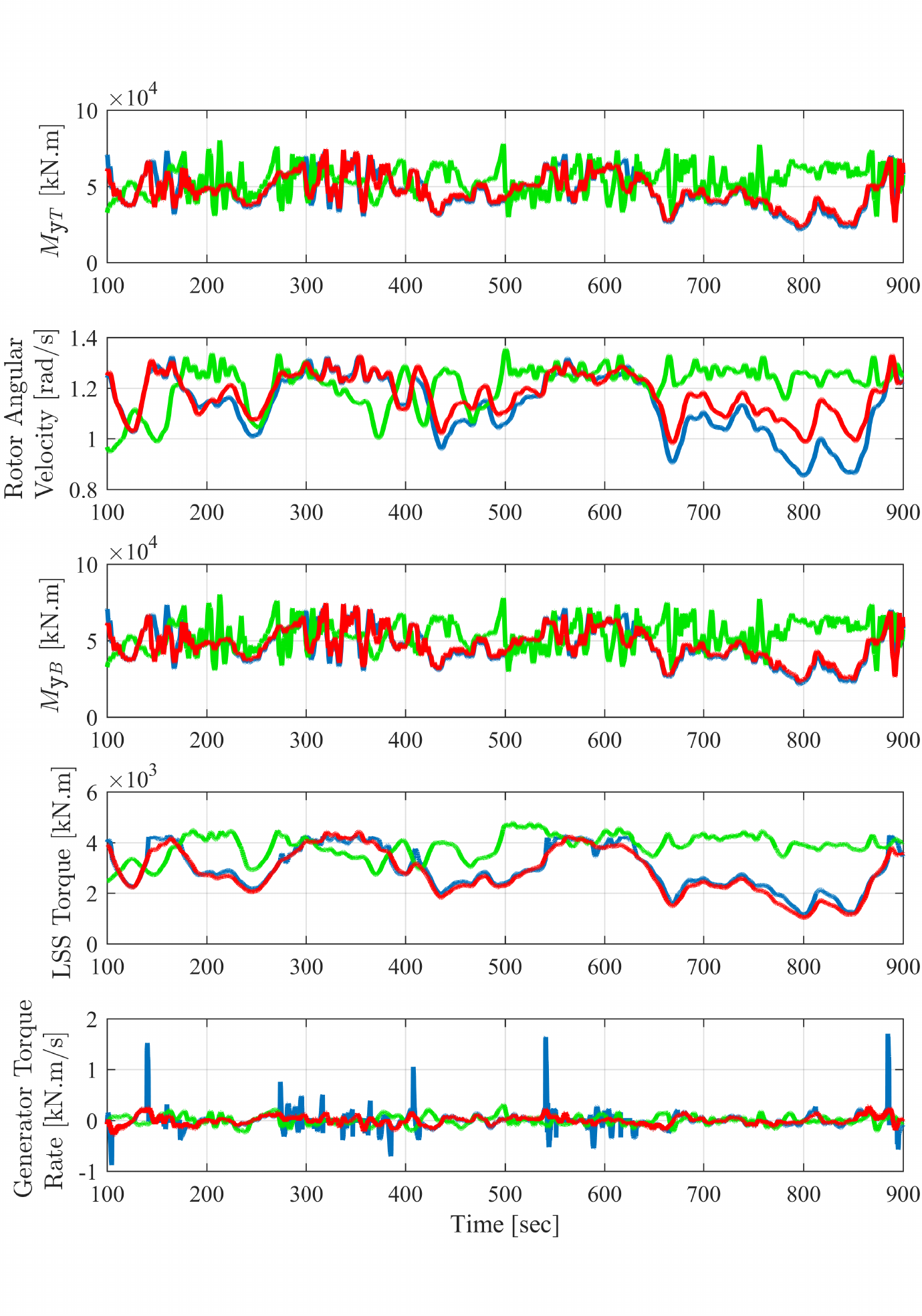}
		\caption{Illustrative responses for mean wind speed of 9 $m/s$ using  Baseline control (Green), DAC  (Blue) and EOR (Red).}
		\label{fig:Region2_TimeRecords}		
	\end{figure}
		
	\begin{figure}[t]
		\centering
		\includegraphics[width=9cm]{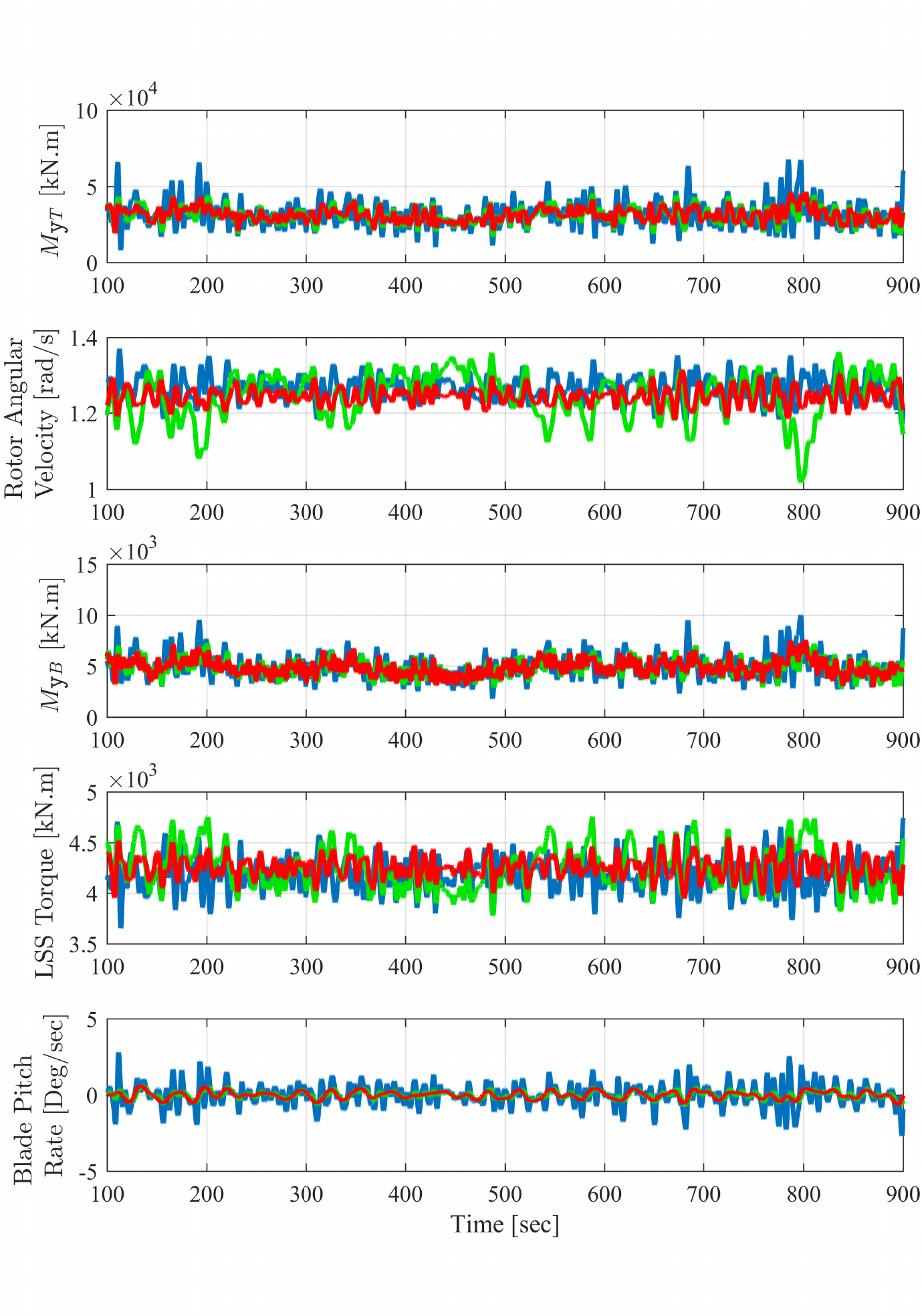}
		\caption{Illustrative responses for mean wind speed of 18 $ m/s$ using  Baseline control (Green), DAC  (Blue) and EOR (Red).}
		\label{fig:Region3_TimeRecords}		
	\end{figure}

Fatigue load DELs, standard deviation of rotor speed and generated power are shown for a range of wind speeds in both operating regions.  A  colour convention  is used to represent the results for different controllers throughout this section as follows: Green represents outputs from a  Baseline controller, Blue represents the DAC and EOR is shown by Red.

Figures \ref{fig:Region2_TimeRecords}	and \ref{fig:Region3_TimeRecords} provide some illustrative time-domain  comparisons. They \cyan{show} 800 seconds of turbine response data under	the three controllers for class-A turbulent wind	signals  of mean  speeds 8 and 18 $m/s$.   The responses from the EOR controller exhibit smaller fluctuations than the two alternative controllers, particularly in the LSS torque  variations. The control input graphs for the  generator torque (Region 2) and  blade  pitch  rate (Region 3)  reveal  that  EOR exerts considerably smoother control  actuation than both Baseline and DAC.		

\subsection{Damage Equivalent Load Controller Performance Comparisons}

%%%%%%%%%%%%%% DEL Graphs
\begin{figure}[!th]
		\begin{subfigure}{.5\textwidth}
		\centering
		\includegraphics[trim={19mm 0 4mm 0},clip, width=\textwidth]{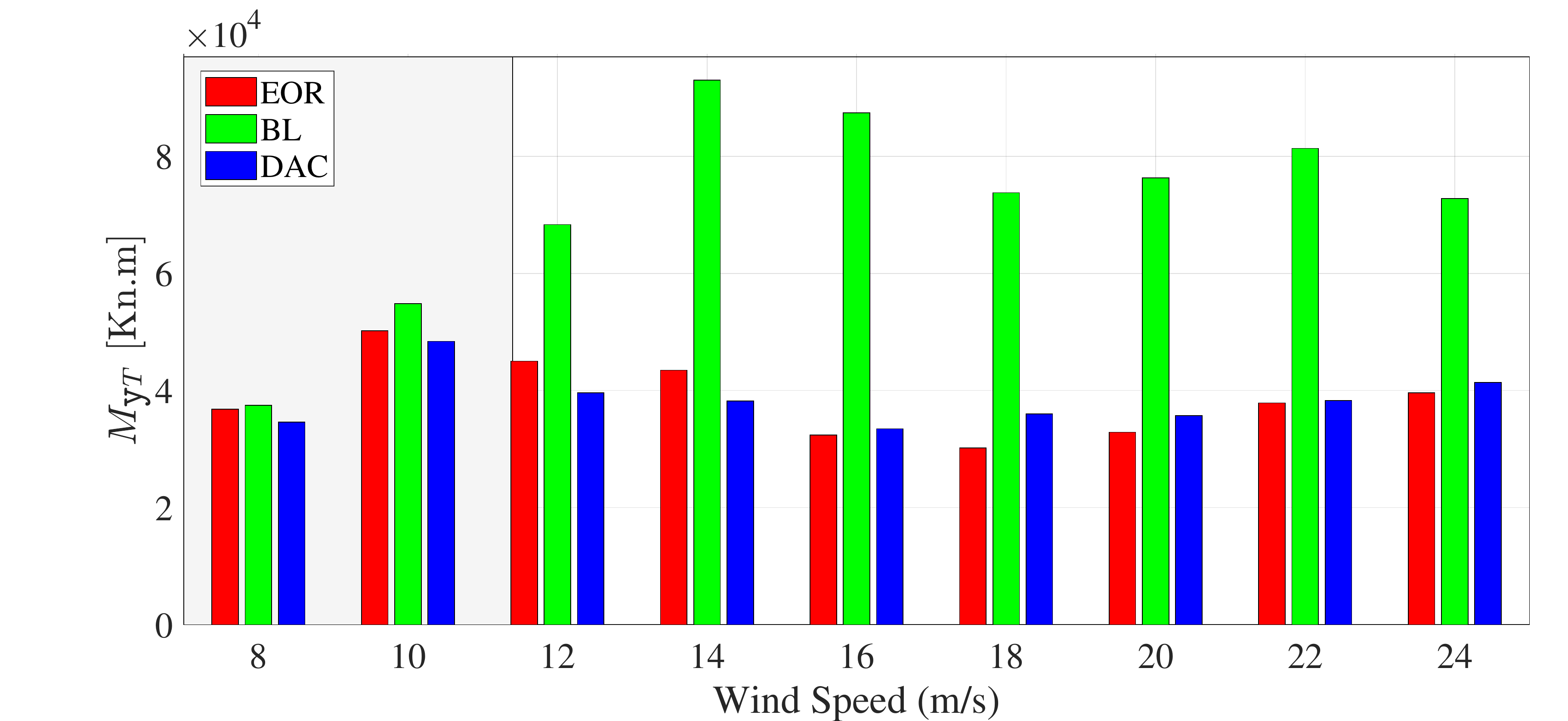}
		\caption{ Tower root fore-aft bending moment DEL }
		\label{fig:MyT_Graph}
	\end{subfigure}
	\begin{subfigure}{.5\textwidth}
		\centering
		\includegraphics[trim={19mm 0 4mm 0},clip, width=\textwidth]{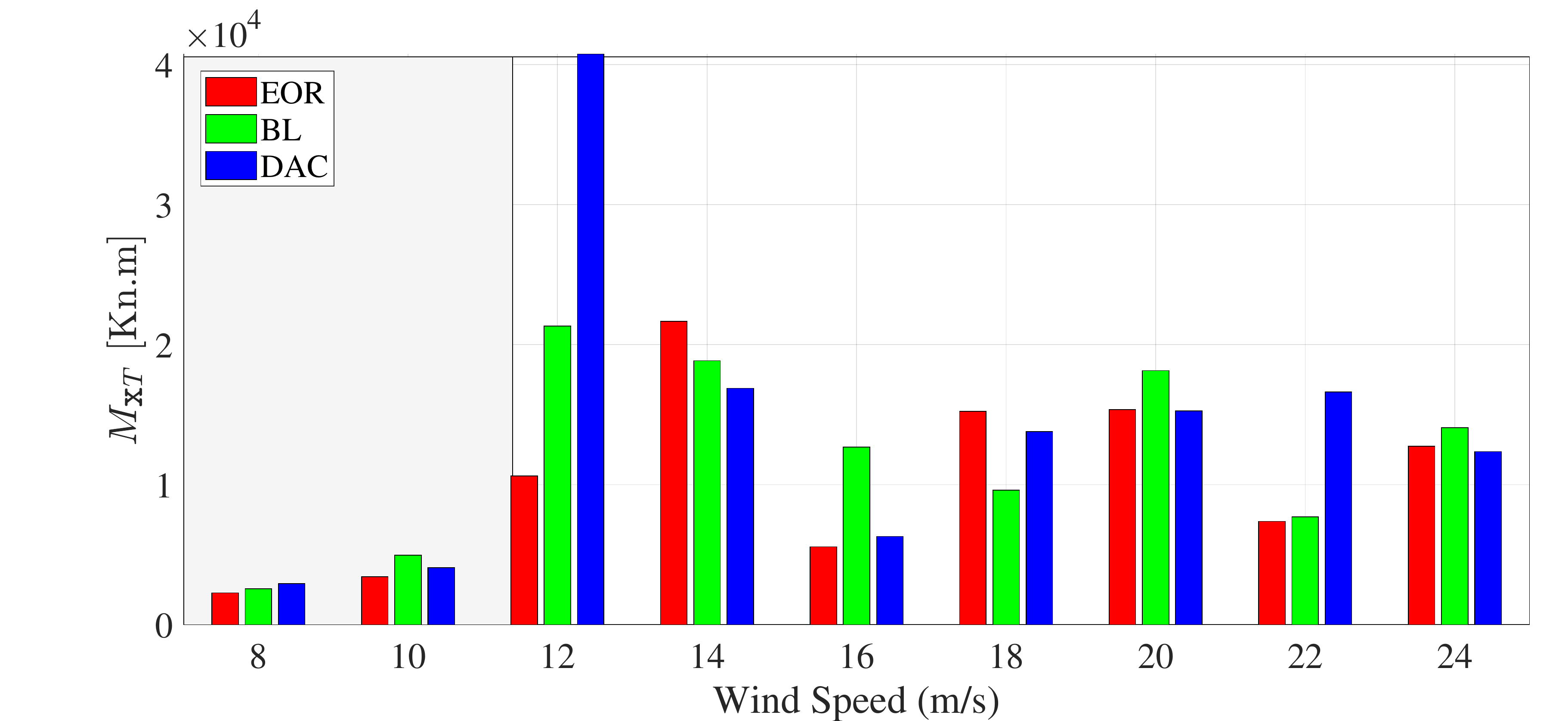}
		\caption{Tower root side to side bending moment DEL}
		\label{fig:MxT_Graph}
	\end{subfigure} \\
	
	\begin{subfigure}{.5\textwidth}
		\centering
		\includegraphics[trim={19mm 0 4mm 0},clip, width=\textwidth]{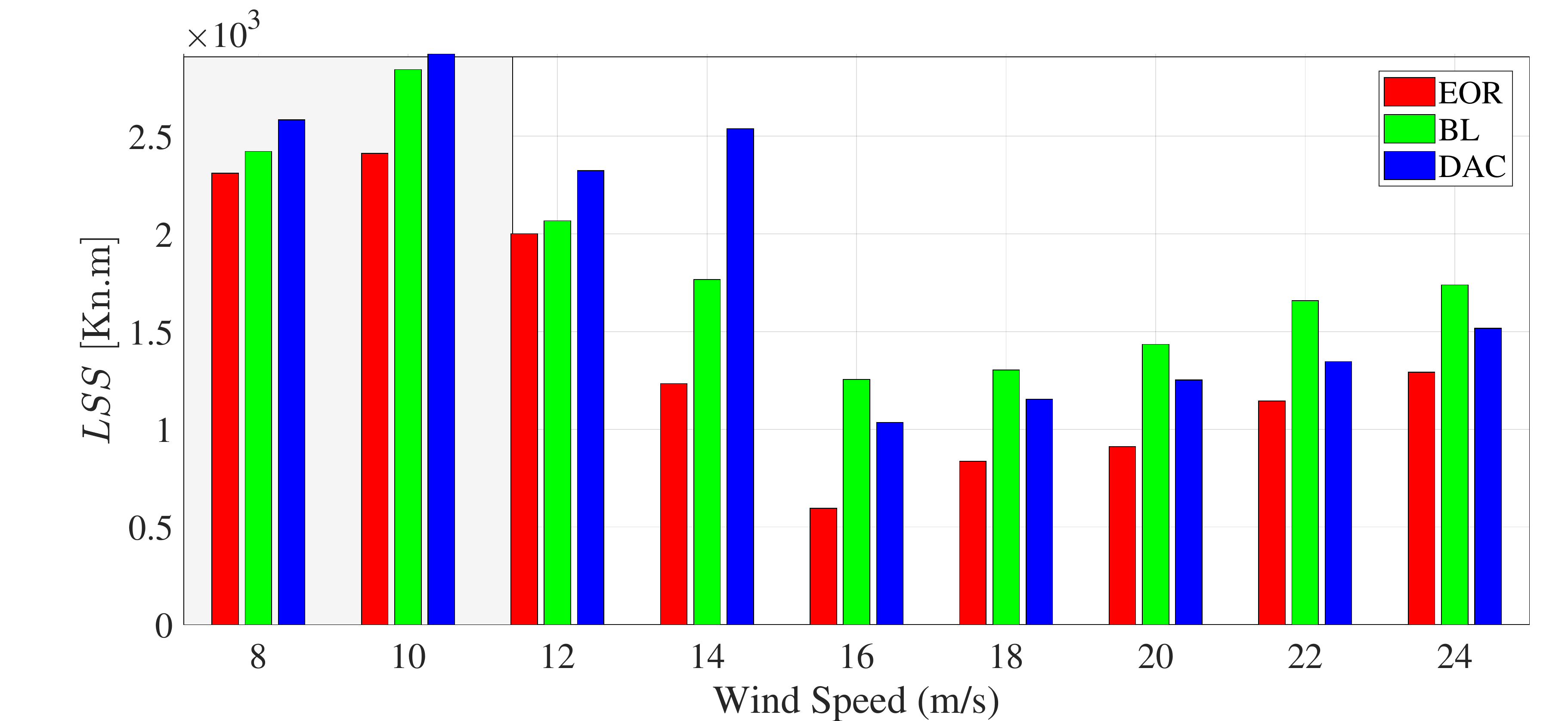}
		\caption{Low speed shaft torsion DEL }
		\label{fig:LSS_Graph}
	\end{subfigure}
	\begin{subfigure}{.5\textwidth}
		\centering
		\includegraphics[trim={19mm 0 4mm 0},clip, width=\textwidth]{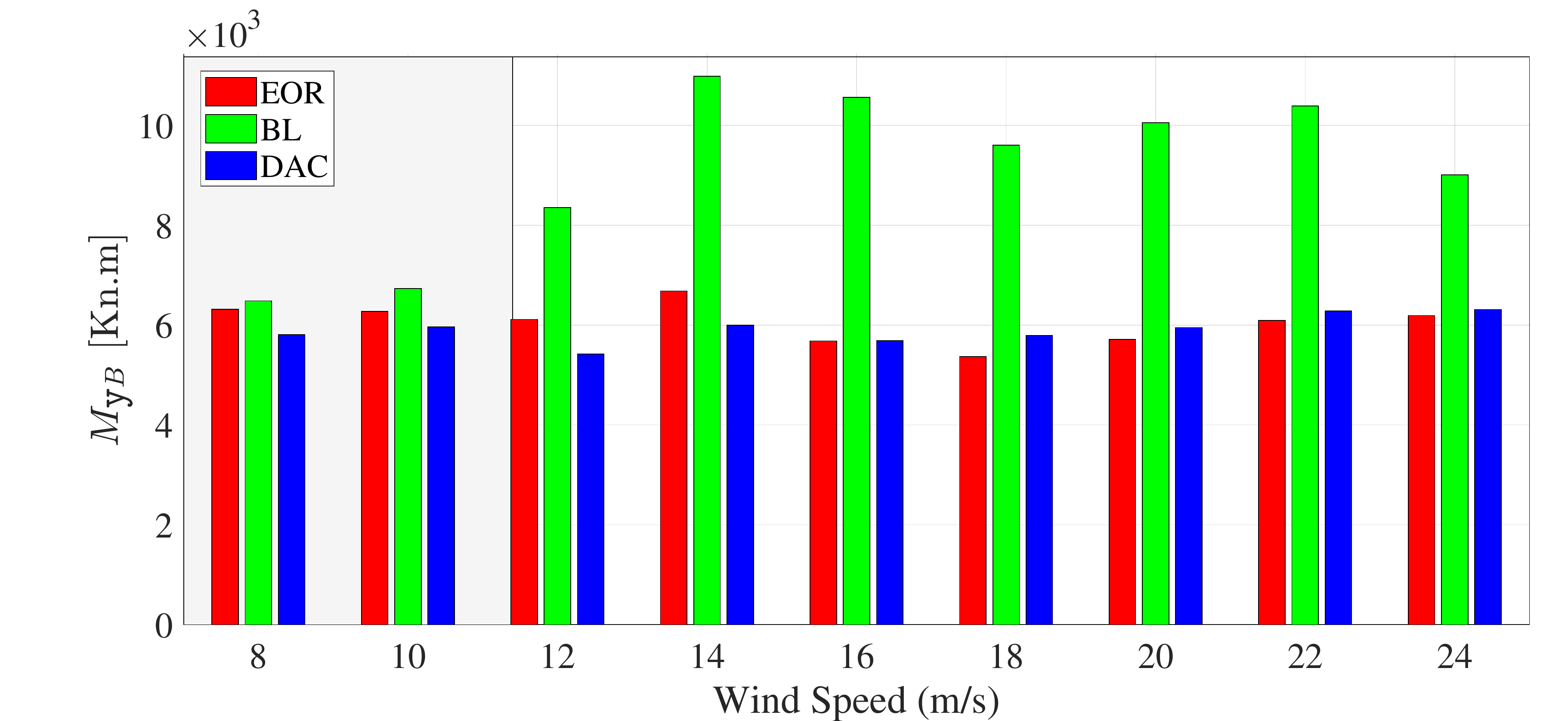}
		\caption{Blade root flap-wise bending moment DEL}
		\label{fig:MyB_Graph}
	\end{subfigure} \\	
	\caption{Damage equivalent loads for 
	 Class-A turbulent winds with mean  wind speeds from  8 $m/s$ to  24 $m/s$.}
	 \label{fig:Reg3_DELs}
\end{figure}

Figures \ref{fig:MyT_Graph} to \ref{fig:MyB_Graph} show fatigue loads for mean wind speeds ranging between  8 $m/s$ to  24 $m/s$.  Results for Region 2 are distinguished with a gray background. In figure \ref{fig:MyT_Graph}, tower root fore-aft bending moment shows that DAC and EOR both improve considerably  over  exo in reducing this load. For tower root  side-to-side bending moment,  Figure \ref{fig:MxT_Graph} does not indicate a consistent improvement for EOR or DAC over  Baseline, apart from the transition region near  12 $m/s$  where  EOR improved greatly over both Baseline and DAC.
In Figure \ref{fig:LSS_Graph}, low speed shaft torque performance is  illustrated,  showing EOR outperforms both alternative controllers across all  wind speeds.  For blade root  flap-wise moments (Figure \ref{fig:MyB_Graph}),  EOR and DAC were  consistently better than Baseline in reducing the blade loads.

\begin{figure}[t]
\centering
	\includegraphics[trim={0 11mm 0mm 0},clip,width=.49\textwidth]{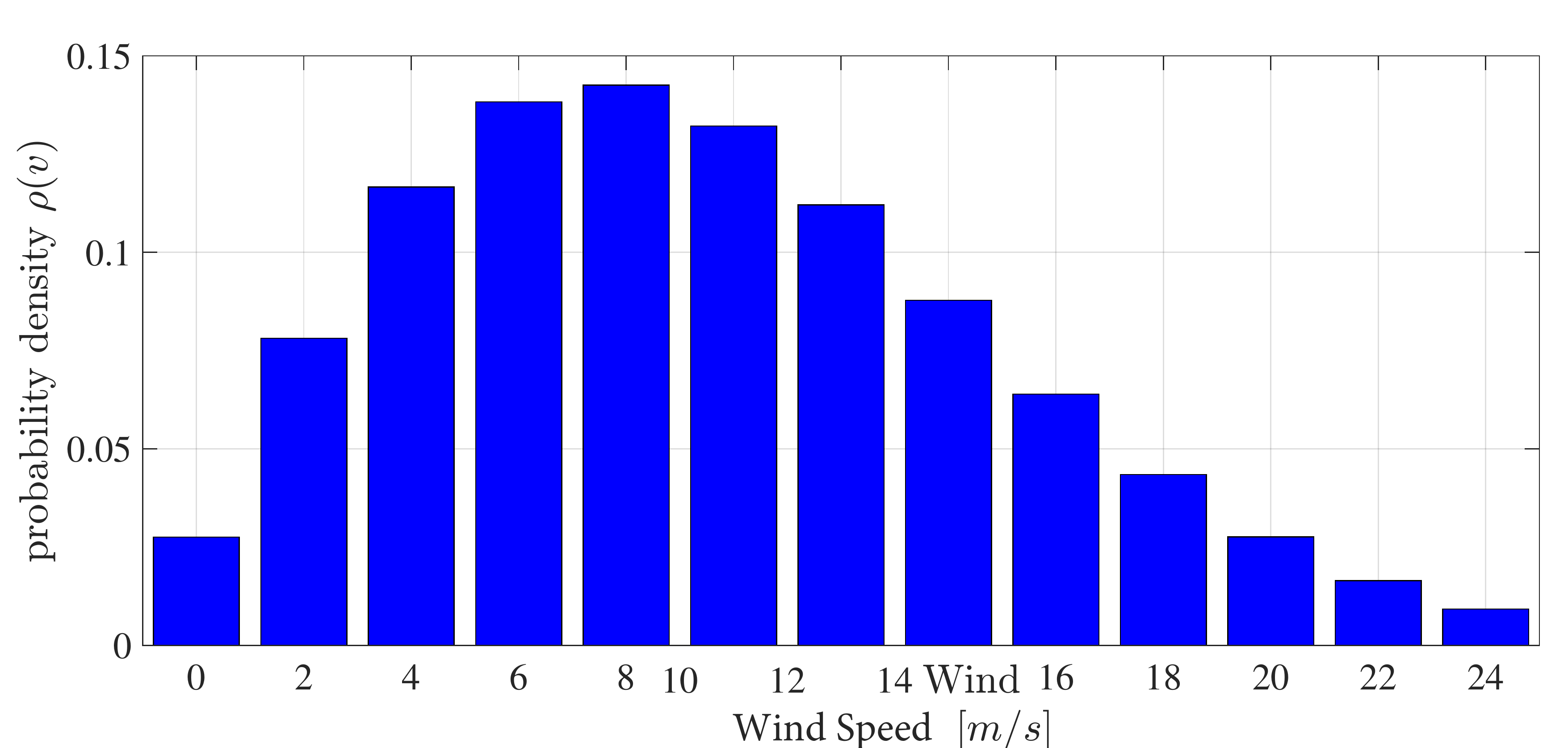}
	\put(-137,-5){\tiny wind speed ($m/s$)}		
	\caption{Weibull distribution of wind speed variation from measured data  at a height of 102 $ m$  in Bremerhaven  \cite{schlipf2016lidar}.}
	\label{fig:Weibull}
\end{figure}

In order to compare lifetime DELs under each controller, the  DELs applicable at each mean wind speed must be averaged across the  operating  range,  with weighting according to  the relative frequency of each mean wind speed.
%Typically, the Weibull distribution is used to determine the frequency of occurrence of each mean  wind speed.
Figure \ref{fig:Weibull} represents a sample Weibull distribution of wind speed variation from measured data at the height of 102 $m$ in Bremerhaven, Germany,  recorded during the winter of 2009 \cite{schlipf2016lidar}.
We  have used this distribution  to weight  the  performance results shown in Figure  \ref{fig:MyT_Graph} to \ref{fig:MyB_Graph},  and the calculated lifetime  values  are shown in the first three rows of  Table \ref{tab:results_R23_A}. The last two rows of this table show  the percentage of improvements of EOR and DAC against the Baseline controller.  Positive numbers indicate  improvement relative to Baseline,  while  negative values indicate inferior performance.

Table \ref{tab:results_R23_A} illustrates that, without reducing power generation, both the EOR and DAC controllers have been able to reduce lifetime DEL loads, in  comparison  with Baseline. However, the DAC performance showed deterioration  in  LSS  torque and tower root side-to-side bending moment,  relative to the  Baseline controller.  By contrast, with the exception of $M_{\texttt{x}B}$,  EOR has been able to improve on Baseline for all the DEL metrics  by  margins  of  between $13 \%$ and $41 \%$.

\begin{table}[h]
\caption{Weighted average of DEL and Power results for class A turbulent wind in both regions}
	\label{tab:results_R23_A}	
	\centering
	\begin{tabular}{ccccccc}
		\hline
		DELs \& STD & \thead{ $M_{\texttt{y}T}$ :[kNm]}
		& \thead{$M_{\texttt{x}T}$ :[kNm]}		
		& \thead{$M_{\texttt{y}B}$ :[kNm]}
		& \thead{$M_{\texttt{x}B}$ :[kNm]}
		& \thead{$LSS$ :[kNm]}
		& \thead{$P_{mean}$ :[MW]}
		\\			
		\hline	\hline
		\thead{Baseline}		 & 8.86E+04 & 2.50E+04  & 1.05E+04  & 1.96+04  & 2.59E+04 & 3.75 	\\
		\thead{EOR} 			 & 5.20E+04 & 1.98E+04  & 6.68E+03  & 1.95E+04 & 2.28E+04 & 3.74	\\
				\thead{DAC}				 & 5.72E+04 & 3.66E+04  & 6.07E+03  & 1.95E+04 & 2.72E+04 & 3.74	\\
		\thead{ EOR \red{\textit{cf.}}  BL \%}       & 41.3    &  20.8    &  33.5    & 0.5      & 13.1    &  $\sim$ 0 \\		
		\thead{ DAC  \red{\textit{cf.}} BL \%}       & 35.4    & -46.4    &  39.6    & 0.5      & -5.02    &  $\sim$ 0 \\
	\end{tabular}
\end{table}

\begin{figure}[h]
		\begin{subfigure}{.5\textwidth}
		\centering
		\includegraphics[trim={8mm 0 4mm 0},clip, width=\textwidth]{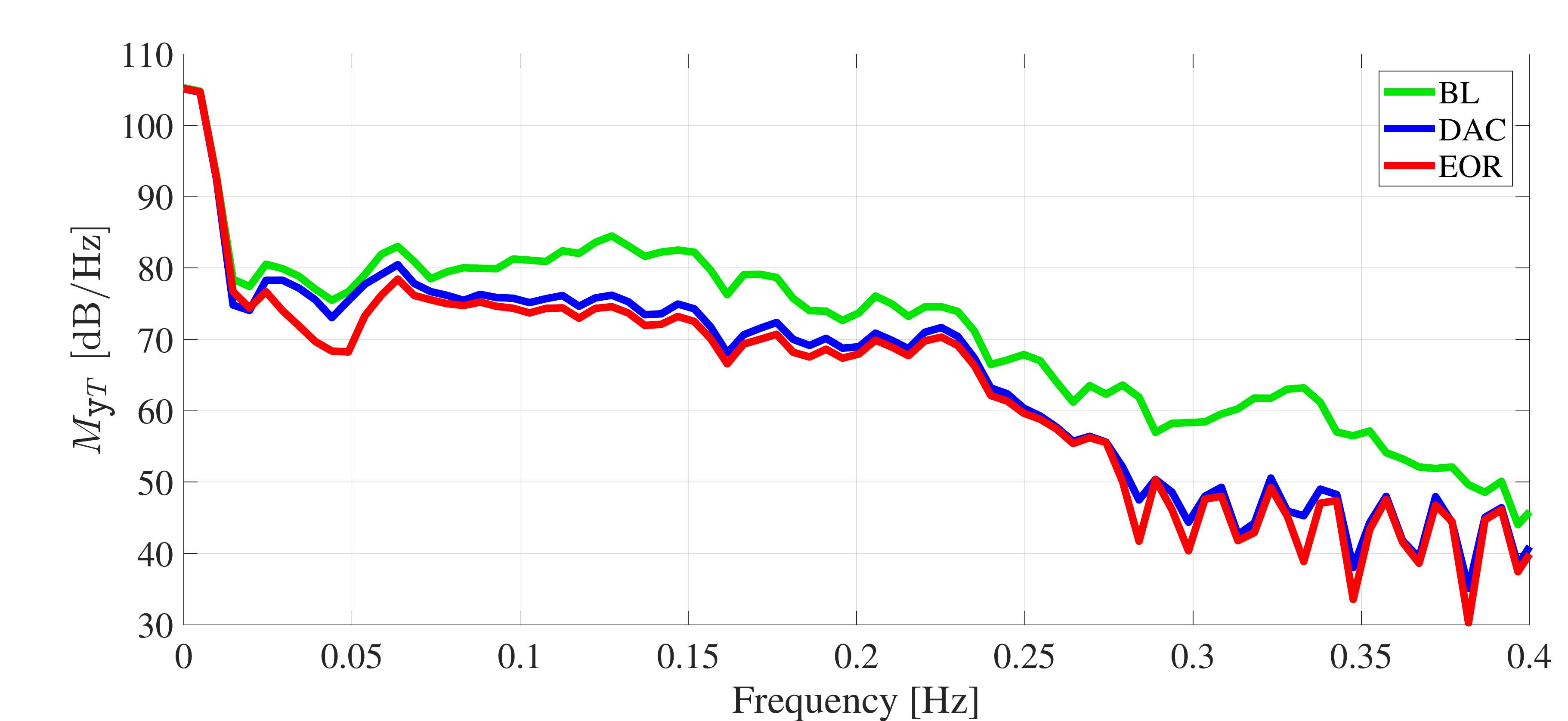}
		\caption{PSD of Tower root fore-aft bending moment $M_{\texttt{y}T}$ }
		\label{fig:PSD_MyT}
	\end{subfigure}
	\begin{subfigure}{.5\textwidth}
		\centering
		\includegraphics[trim={8mm 0 4mm 0},clip, width=\textwidth]{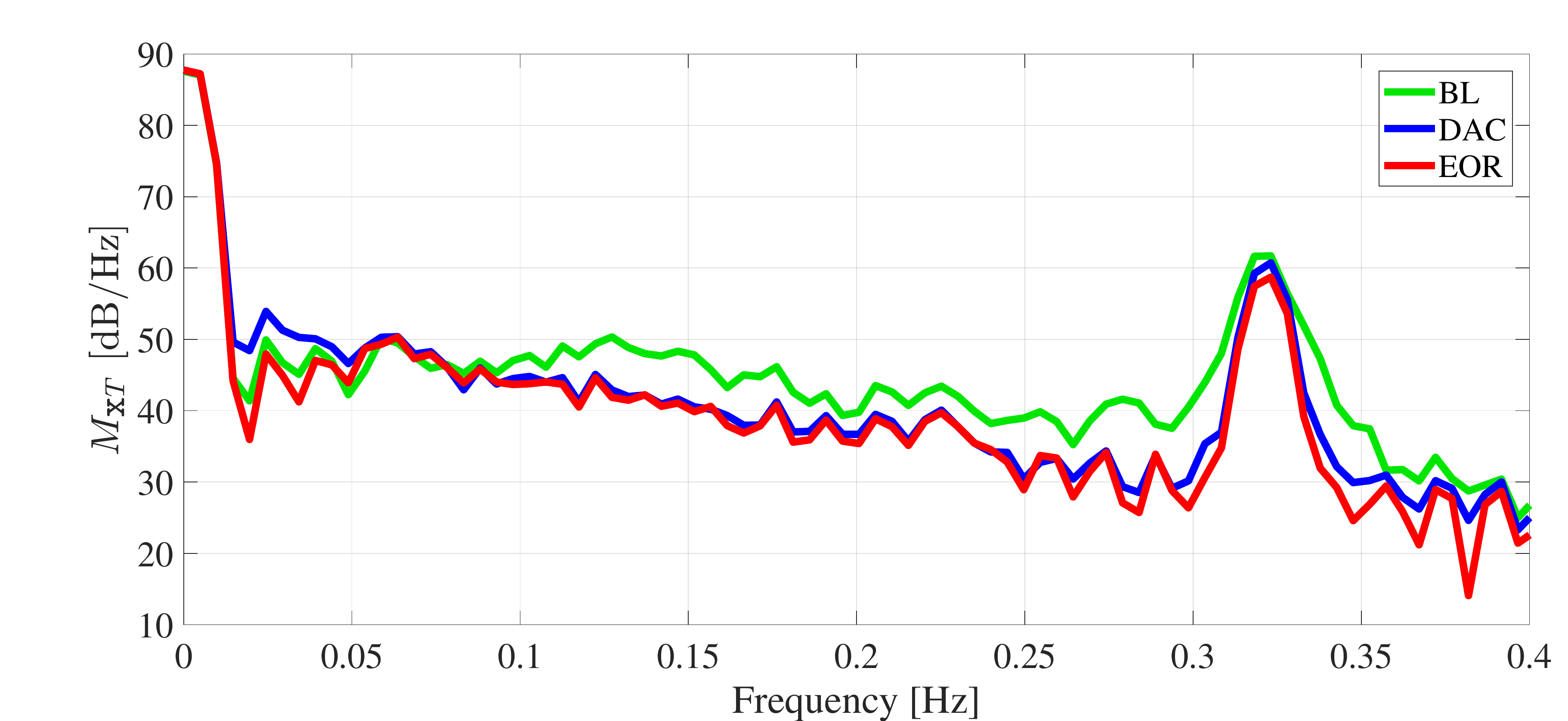}
		\caption{PSD of Tower root side to side bending moment $M_{\texttt{x}T}$}
		\label{fig:PSD_MxT}
	\end{subfigure} \\
	
	\begin{subfigure}{.5\textwidth}
		\centering
		\includegraphics[trim={8mm 0 4mm 0},clip, width=\textwidth]{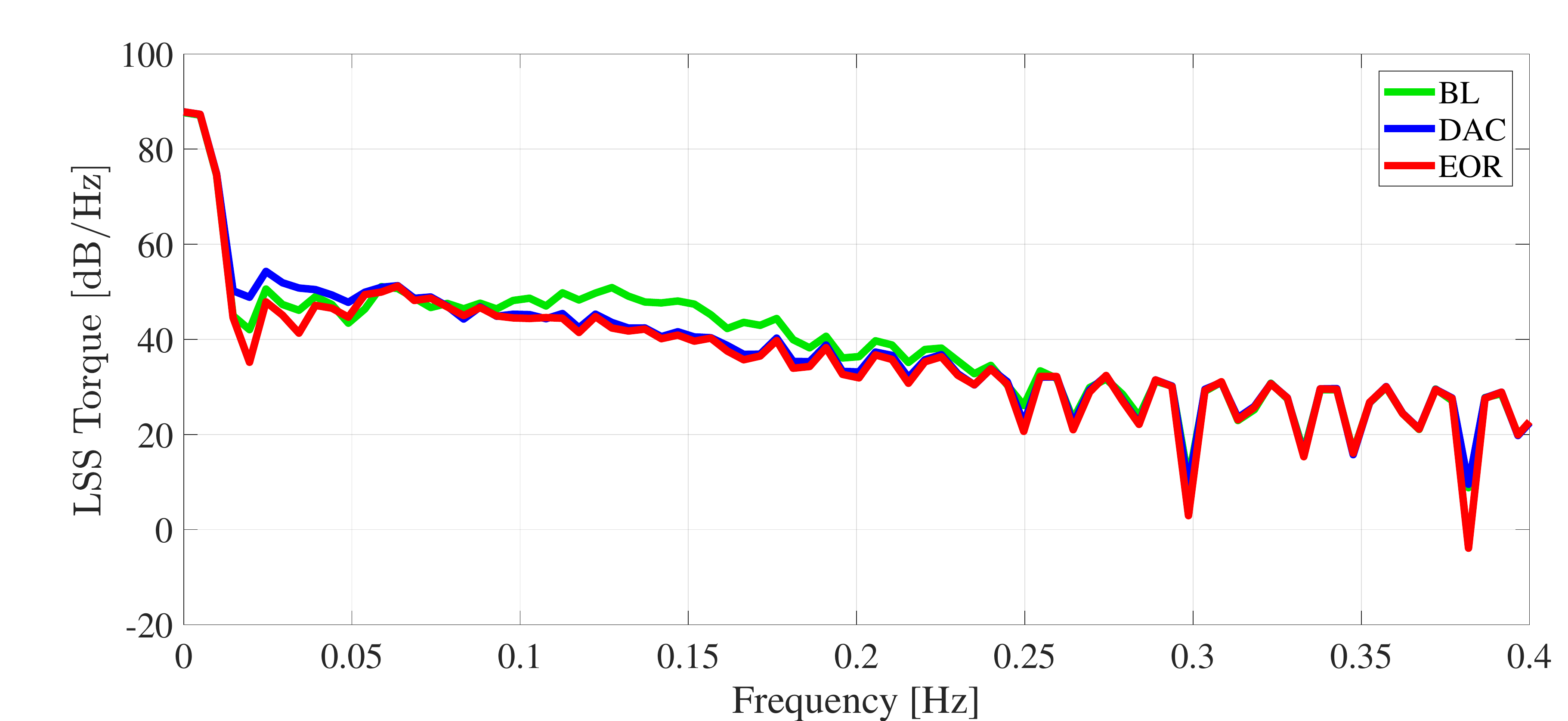}
		\caption{PSD of Low speed shaft torsion $LSS$ }
		\label{fig:PSD_LSS}
	\end{subfigure}
	\begin{subfigure}{.5\textwidth}
		\centering
		\includegraphics[trim={8mm 0 4mm 0},clip, width=\textwidth]{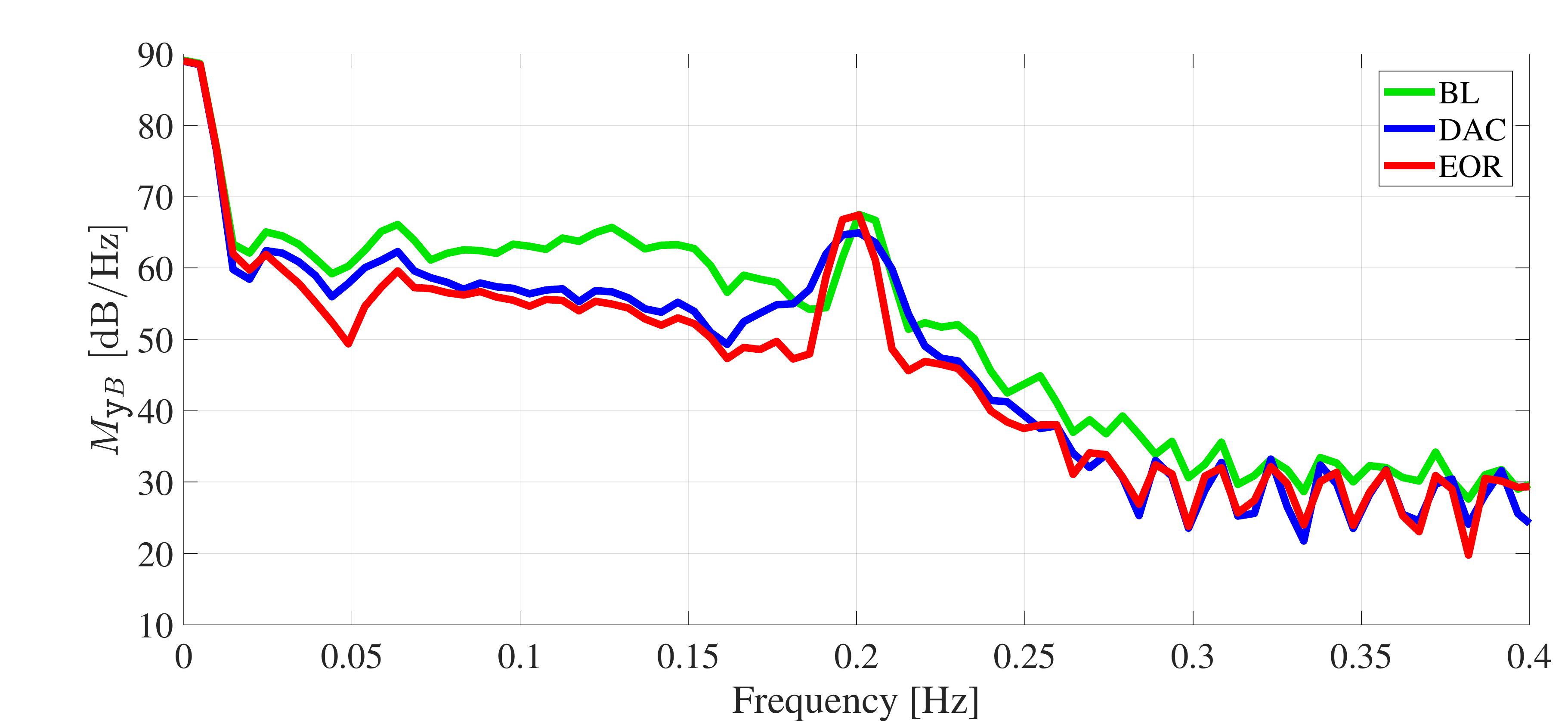}
		\caption{PSD of Blade root flap-wise bending moment $M_{\texttt{y}B}$}
		\label{fig:PSD_MyB}
	\end{subfigure} \\
		\caption{Power Spectral Density Graphs of \red{Loads}  at the wind speed 20 $m/s$.} \label{fig:PSDset}
\end{figure}

Figure \ref{fig:PSDset} shows power spectral densities for $M_{\texttt{y}T}$, $M_{\texttt{x}T}$,   $LSS$  and $M_{\texttt{y}B}$
%, as well as  $\dot{\theta}$ and generated power,
 for a  mean wind speed of 20 $m/s$. In very low frequencies all controllers show similar spectral content,  but above 0.02 $Hz$, EOR has  superior  attenuation.
As can be seen in Figure \ref{fig:PSD_MyT} for the tower root fore-aft bending moment $M_{\texttt{y}T}$, EOR shows the most reduction around $0.05 Hz$ while for higher frequencies the EOR performance is comparable to  DAC.
For the tower root side-to-side bending moment $M_{\texttt{x}T}$ shown in  Figure \ref{fig:PSD_MxT}, similar improvements can be seen for EOR. The excitation around 0.32 $Hz$ represents the first  tower root side-to-side natural frequency \cite{jonkman2009definition}. Since the tower natural  frequency is  not considered in the reduced model \eqref{eq:torsion}-\eqref{eq:pitchservo}, this frequency is not attenuated by any of the controllers.
For low speed shaft  torque in Figure \ref{fig:PSD_LSS}, DAC falls short of the other controllers below 0.05 $Hz$ while EOR maintains better attenuation across the spectrum.

Blade root flap-wise moments  $M_{\texttt{y}B}$ in Figure \ref{fig:PSD_MyB} show a peak at 0.2 $Hz$ which is the 1P frequency (1 times the rotor frequency) for all three controllers.  EOR and DAC show  similar improvement over Baseline in the other parts of the spectrum.

%The most noticeable performance difference happens on the pitch rate illustrated by Figure \ref{fig:PSD_dTheta}. EOR  substantially reduces the high frequency commands on the pitch actuator at the cost of negligible increase in lower frequencies indicating smoother control actuation. Finally, EOR shows slightly better performance on the smoothness of generated power (Figure\ref{fig:PSD_Power}).
%The DELs presented in Table \ref{tab:results_R23_A} are a weighted average of the whole operating range, and are consistent with the PSD analysis of a $20$ m/s reference wind speed signal.

%%%%%%%%%%%%%%%%%%%%%%%%%%%%%%%%%%%%%%%%%%%%%%

\subsection{Power Generation Controller Performance Comparison}

\begin{figure}[h]
%	\label{fig:Reg3_PWR}
	\begin{subfigure}{.5\textwidth}
		\centering
		\includegraphics[trim={10mm 0 4mm 0},clip, width=\textwidth]{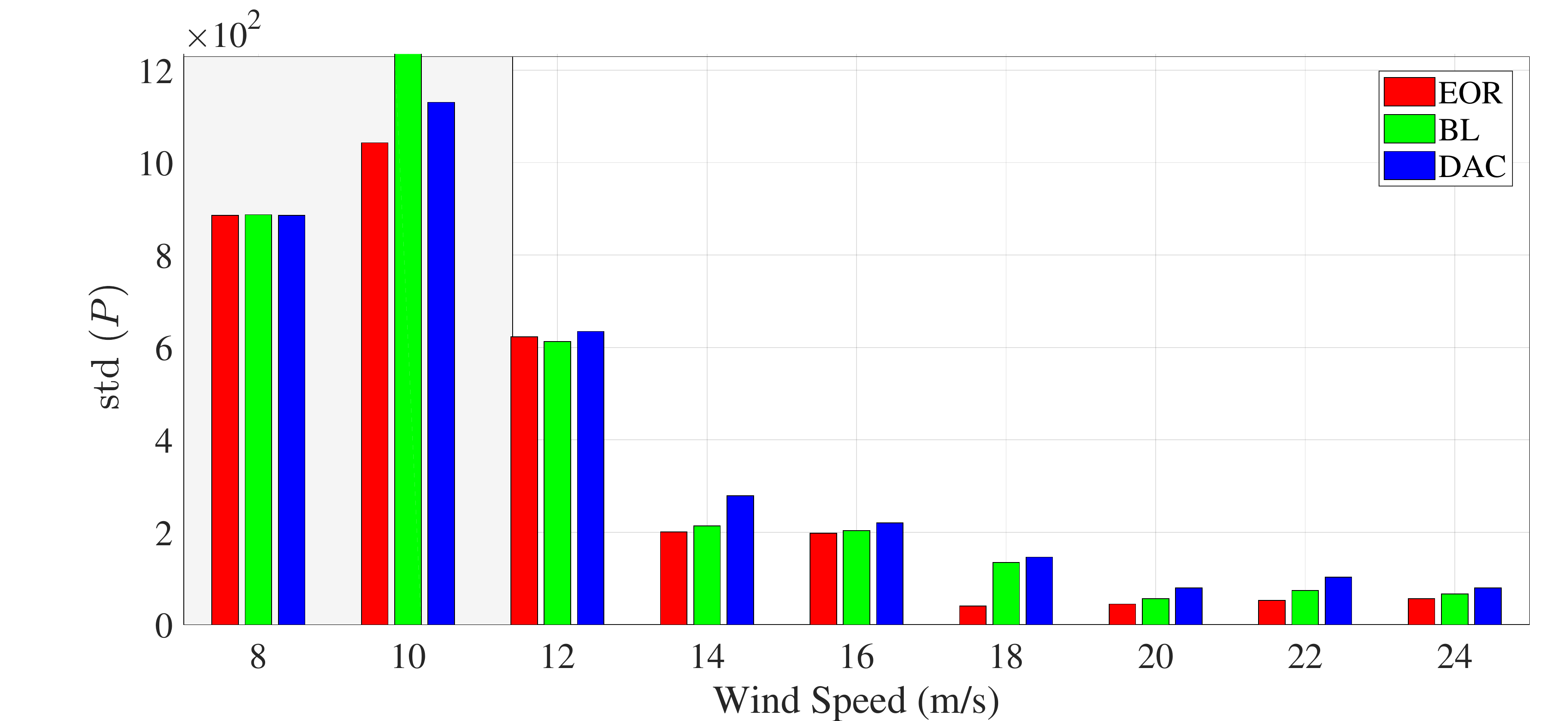}
		\caption{Standard deviation of electric power generation}
		\label{fig:Power_Graph}
	\end{subfigure}
	\begin{subfigure}{.5\textwidth}
		\centering
		\includegraphics[trim={10mm 0 4mm 0},clip, width=\textwidth]{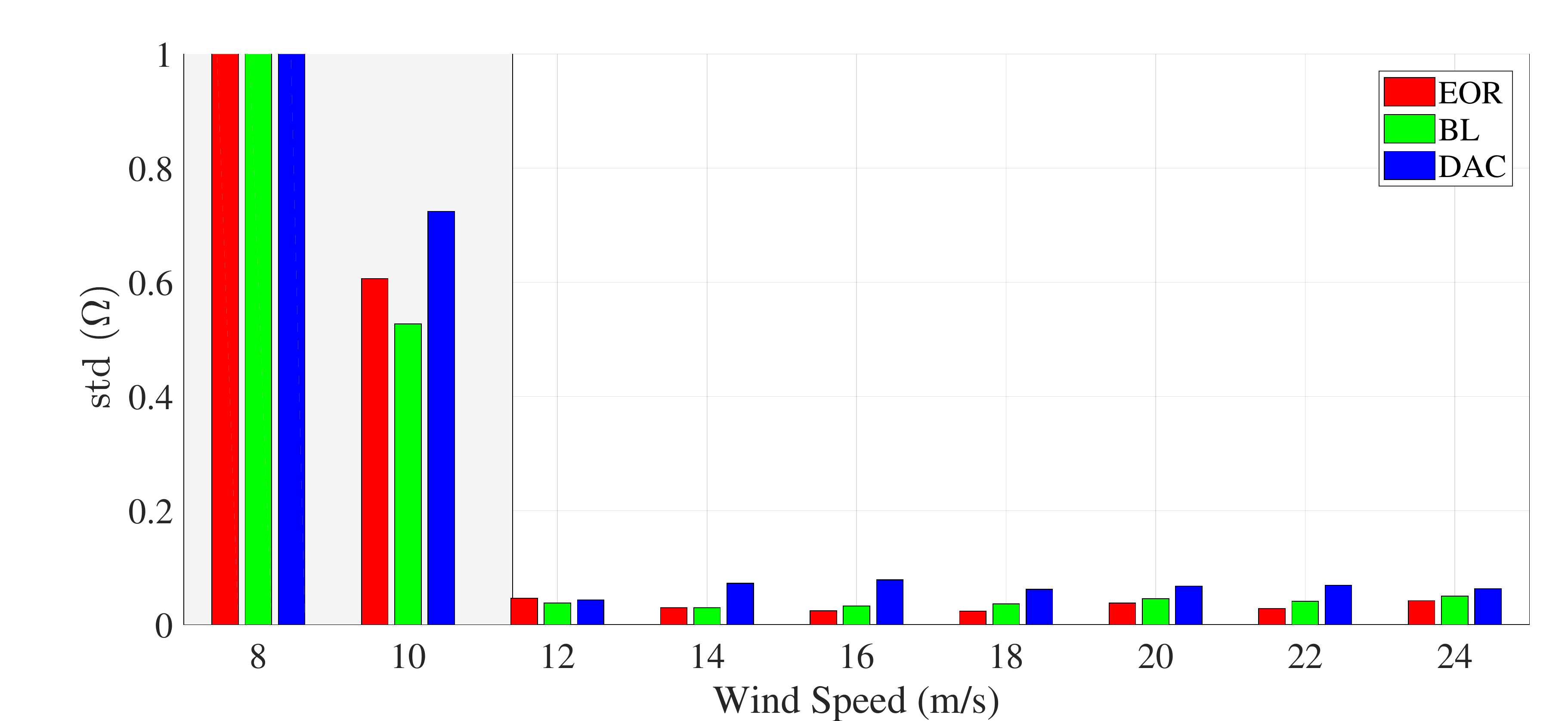}
		\caption{Standard deviation of rotor speed }
		\label{fig:Omega_Graph}
	\end{subfigure}
		\caption{Standard deviation of power generation  and  rotor speed for Region 3 mean  wind speeds} \label{fig:PSDset} 	
\end{figure}

To  compare the power generation performance of  the  three controllers, we  distinguish between Region  2 and  Region  3  performance.
Figures  \ref{fig:Power_Graph} and  \ref{fig:Omega_Graph}  illustrate the standard deviations  of  rotor speed and generated power. Reducing these standard deviations  implies less variation in these variables, indicating that the controller gives better  performance in maintaining the rotor speeds and rated power at their rated values.
These two graphs only contain Region 3 wind speeds, as these objectives only apply in  Region 3 operation. Here EOR again outperforms both controllers, with  the DAC controller giving  significantly worse performance.

\begin{table}[h]
	\caption{Power Generation Performance}
	\label{tab:results_R2_A}
	\centering
	\begin{tabular}{cccccc}
		\hline
		& \multicolumn{2}{c}{Region 2 (9 m/sec)} & \multicolumn{3}{c}{Region 3}\\
		& \multicolumn{1}{c} \thead{$std(\lambda)$}
		& \thead{$P$ :[MW]}
		& \thead{$std(\Omega_r)$}
		& \thead{$std(P)$}
		& \thead{$P$ :[MW]}
		\\			
		\hline	\hline
		
		\thead{EOR} 			 & 0.534   & 2.15 & 0.034 & 153 & 4.71 \\
		\thead{Baseline}		 & 0.409   & 2.15 & 0.041 & 171 & 4.71 \\
		\thead{DAC}				 & 0.641   & 2.15 & 0.069 & 211 & 4.68 \\
		%\hline
		\thead{ EOR \red{\textit{cf.}} BL \%} 		 & -30.6  & $\sim$ 0&  17.9   &  10.5  & $\sim$ 0\\		
		\thead{ DAC \red{\textit{cf.}} BL \%} 		 & -56.7  & $\sim$ 0& -67.1   & -23.4  &  -0.6 \\
	\end{tabular} 		
\end{table}

Table \ref{tab:results_R2_A} summarizes the results of the controller performance for  power generation. In Region  2,  we  have chosen a  single wind speed  of  9 $m/s$  to  ensure the wind speed signal mostly remains above
7.85 $m/s$.   For wind speeds  below  this level, the Baseline controller is not designed to track  the  optimal TSR,  and  this would invalidate the controller energy harvesting performance comparison.

 The first column of the table shows that  in Region 2,  both EOR and DAC have considerably higher standard deviation in their  TSR $\lambda$, indicating less rigid control of the rotor speed.
 While this  might  be expected to indicate a  failure to achieve the optimal TSR for  energy  generation,  column  2 of the table  indicates    negligible differences in the energy harvested by the three controllers.  Thus the load reductions  observed  in Figures \ref{fig:MyT_Graph}  to  \ref{fig:MyB_Graph} have been achieved  without sacrificing power generation.

 The averaged  results for the  standard deviations  $std(\Omega_r)$ and $std(P)$ are
 shown in columns 3 and 4 of  Table \ref{tab:results_R2_A},  with wind speeds weighted according to the Region  3   wind speeds in the Weibull distribution of  Figure \ref{fig:Weibull}.    The EOR controller achieved a substantial improvement in both rotor  speed and power regulation by $17.9 \%$ and $10.5 \%$, relative to  Baseline. Conversely,  DAC  suffered a performance degradation of $67.1 \%$ and $23.4 \%$,  relative to  Baseline.  Column  5 of Table   \ref{tab:results_R2_A}  show that    EOR generated the same amount of power as Baseline, while DAC showed some very slight reduction in  power generation.

Reducing rotor speed standard deviation in Region 3 reduces the likelihood that the rotor speed will  violate the safe operational limits on  $\Omega_r$. This reduces the chances of turbine failure  and  increases the turbine's operational  availability.

%%%%%%%%%%%%%%%%%%%%%%%%%%%%%%%%%%%%%%%%%%%%%%%%%%%%%%%%%%%%%%%%%
\subsection{Pitch Actuation and Command Torque Rate }		
\begin{figure}[h]
			\centering
			\includegraphics[trim={24mm 10mm 0mm 0},clip,width=9cm]{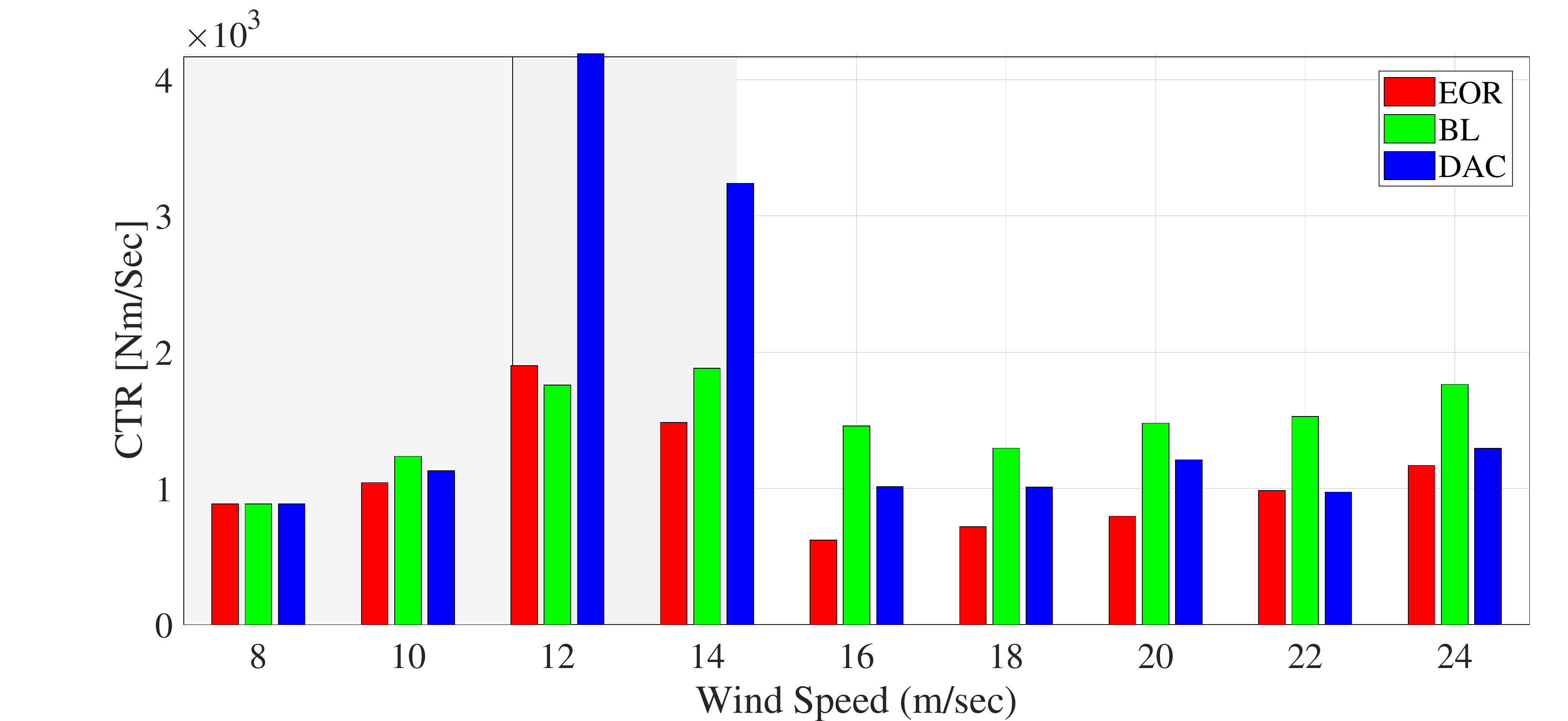}
			\put(-147,-5){\scriptsize	 wind speed ($m/s$)}	
			\caption{Command Torque Rate for Class-A turbulent winds with mean wind speeds from  8 to  24 $m/s$. }
			\label{fig:CTR_Graph}
		\end{figure}

			The command torque rate (CTR) of the generator torque  is  given by
				\begin{equation}
				CTR = \sqrt{\frac{1}{T} \int_{0}^{T} \left(\frac{d M_g(t)}{dt}\right)^2 dt}
				\label{eq:CTR}
			\end{equation}
	The CTR measures  changes to  the generator torque set point during the 60 minute simulation  period.
	From the system  dynamic equations  \eqref{eq:genspeed}-\eqref{eq:torsion},  we   observe that changes in  $M_g$ cause changes in the drive train torsion,  and these variations  are associated with fatigue on the drive shaft. Therefore, it is desirable for a controller to achieve its control objectives with reduced torque actuation.
			Figure 	\ref{fig:CTR_Graph} shows the CTR for Region  2 and 3   wind speeds.

		The results show that EOR has the lowest CTR  in all wind speeds except 12 $m/s$, indicating smoother torque control  in both operating regions. DAC has  lower CTR compared to Baseline except  at wind speeds close to the transition region (12 to 14  $ m/s$).		
		\begin{figure}
			\centering
			\includegraphics[trim={20mm 10mm 0mm 0},clip,width=9cm]{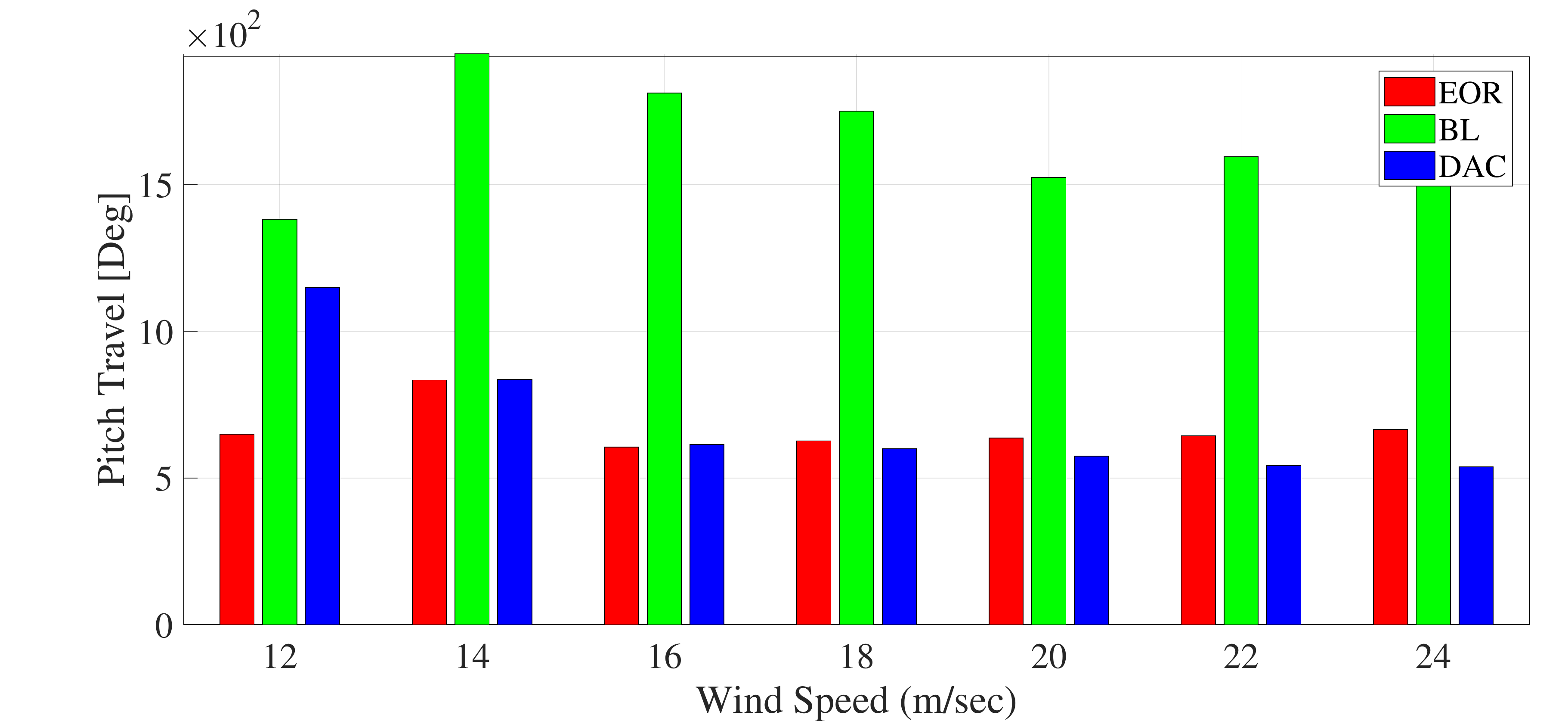}
			\put(-147,-5){\scriptsize	 wind speed ($m/s$)}				
			\caption{Pitch Travel in Region 3 for Class-A turbulent winds with mean wind speeds from  12 $m/s$ to  24 $m/s$. }
			\label{fig:PitchTravel_Graph}
		\end{figure}		
			The pitch travel  (PT) of the blade angle over the  one-hour simulation period  is defined to  be 		
			\begin{equation}
				PT = \int_{0}^{T} \left|\frac{d \theta}{d t} \right| dt
				\label{eq:PT}
			\end{equation}
	Reducing  pitch travel reduces wear and tear on the bearings of the pitch mechanism.	%Additionally, pitch travel indirectly affects the tower root fore-aft and blade flap-wise fatigue loads.  Rapid changes in blade pitch angle are correlated with rapid  changes in aerodynamic thrust. Such variations vary the force of the wind on  the tower,  contributing to tower fatigue and possibly also blade  fatigue. 
	Figure \ref{fig:PitchTravel_Graph} shows  the  pitch travel for  Region  3 wind speeds, where pitch actuation is active.

		We observe that EOR and DAC have significantly lower pitch travel compared to Baseline. In higher than 16 m/s wind speeds,  EOR has  slightly higher pitch travel than  DAC. This may be  related to the  superior  performance of EOR in controlling the rotor speed and output power at a more constant level,  which \cyan{requires}  greater pitch actuation.
	
%%%%%%%%%%%%%%%%%%%%%%%%%%%%%%%%%%%%%%%%%%%%%%%%%%%%%%%%%%%%%%%%%%%%%%%%%%%%%%%%%%%%%%%%%%%%%%%%%%%%%%%%%%%%%%%%%%%%%%%%%%%%%%%%%%%%%%%
\section{Conclusion and Future Work}
\label{sec:Conc}

We have proposed a new strategy for wind turbine control using the  classical EOR control  methodology. Simulations were performed with the NREL FAST code for the NREL 5 $MW$ wind turbine model, using a broad range of realistic wind signals generated by TurbSim. Wind preview information is assumed to be available from LIDAR measurements. Our results showed the  EOR controller  was able to provide substantial and consistent fatigue load reductions compared to the Baseline controller and DAC,  without loss of generated power,  in both operating regions. Additionally,  its  modest computational  cost   means EOR  can be expected to be suitable for real-time implementation.

In Region 2 maximizing  the energy production   requires \eqref{cpmax} to hold at all  times,  and this requires  the generator torque controller be designed to precisely track the optimal TSR,  and  all other states of the  turbine to precisely track their equilibrium value for all instantaneous effective wind speeds. However, it was  shown in \cite{bossanyi2014wind} that the power coefficient  $C_p$ is a relatively flat curve at its maximum and a  controller designed to  track the optimal TSR may contribute only very slight increases in power production,  while adding a considerable amount of stress on the structure. Similar results were noted in  \cite{wang2013comparison},\cite{wang2014lidar} and \cite{bossanyi2014wind}, where very small  increases in power production came at the cost of substantial increases in DELs.

This situation suggests that proposed improvements in controller design  methods  should  aim to reduce  load fatigues,  without compromising  power harvesting efficiency.  Our results in Region 2 showed that stress on the structure can be considerably reduced by EOR, without compromising energy capture.   Similarly, our results in Region 3 showed that EOR can reduce  variability in the rotor speed and power generation,  again without   compromising  energy production.

Considering the relative  performance of the  EOR and DAC controllers, our performance comparisons showed an EOR controller was able to  provide superior fatigue load reduction to a DAC controller.
%The improved performance of EOR may be attributed to its ability  to  utilise higher derivatives of the wind disturbance signal.
The  feedforward gain used in DAC does not take the wind dynamics into account,  and this  a restricting assumption when dealing with rapidly varying disturbances. Conversely, EOR accommodates  higher-order dynamics when modelling wind disturbances,  by  constructing an exosystem whose  states include derivatives of the disturbance. Therefore, EOR is better able to utilise the wind  information obtained from LIDAR to  develop wind prediction capability,  by utilising time derivatives of the disturbance signal.
 However, our study  does  not  allow for wind evolution  between  the  LIDAR focal point and the blades, nor does it consider  LIDAR measurement noise.  The presence of either   of these factors can  be expected to  reduce the  observed performance improvements of both EOR and DAC over Baseline.

%One of the main possible reasons that DAC does not achieve much of improvements in spite of the advanced control methodology over Baseline controllers is the assumption of constant disturbances.

The order of the exosystem employed within  the  EOR methodology  can be freely  chosen depending on the reliability of the  LIDAR measurements, computational power and the parameter estimation limitations. Therefore, it is possible to design the EOR controller with regard to different control performance objectives.
For example, a higher-order exosystem will yield a more precise description of the wind and may improve disturbance rejection. However, it may also increase the structural loads and actuation efforts.
Hence, selecting the appropriate order for the exosystem involves  a trade-off between increasing the power generation performance and reducing the loads.

%It can also be speculated that the smallest  feasible value for the order of the exosystem depends on the relative degree of the transfer function between the system input and the disturbance input channel. This will be further investigated in future research.

The classical EOR controller design  method described in section  \ref{secEOR}  has been extended  to  accommodate  other control problem frameworks, such  as the {\it robust output regulation  problem},   in  which  the  objective  is to achieve output  regulation in  the  presence of  plant  uncertainty, and  the {\it nonlinear output regulation problem} which considers the problem of regulating the  output of a nonlinear  plant \cite{huang2004nonlinear}. Both of these variations  on  EOR  have the  potential  to further improve turbine performance, by  accommodating
the  plant uncertainty introduced by use of the linearised model (\ref{eq:Sigma}),  or else through the  direct use of the
 nonlinear model in (\ref{eq:torsion})-(\ref{eq:pitchservo})  for the controller design.

Future developments will also consider the performance of the EOR methodology using individual pitch control. It is anticipated this yield further improvements in rotor speed control with reduced fatigue loads,  in comparison with Baseline and DAC.

Another area for future work is to  develop a transition strategy between Regions 2 and 3. For mean wind speeds near the transition wind speed of
 11.4 \  $m/s$, our  analysis showed EOR  gave reduced improvements than  when the  mean wind speeds were not  close to  transition.  For instance, standard deviation of rotor speed $std(\Omega_r)$  at the reference wind speed of 12 $m/s$ is higher under EOR than both Baseline and  DAC,  in contrast with the superior performance of EOR at all other Region 3 wind speeds.  Similarly, the command torque ratio  at  12 $m/s$ for EOR is higher than Baseline which can clearly be associated with  transition region issues. The EOR performance is clearly better than Baseline  at  all other mean wind speeds.
		Therefore, an EOR transition strategy is  needed  to  enable more effective switching from generator  torque control  to  blade pitch control as wind speeds vary  between Region 2 and 3.

%
%Finally, due to the use of exosystems in EOR, another strong potential for development of EOR is to consider effects of wind evolution after the LIDAR measurement point. Exo-systems can give a fair insight to the controller, how the wind would evolve in short term and make the controller to devise preemptive actions for it.

\section*{Acknowledgment}
The authors would like to thank D.  Schlipf and M. Mirzaei for numerous helpful suggestions,  and the anonymous reviewers
for their  detailed and   insightful  comments.

\end{document}